\documentclass[final,1p,times]{elsarticle}

\usepackage{amssymb}

\usepackage{siunitx}
\usepackage{placeins}
\usepackage{booktabs}
\usepackage{subcaption}
\usepackage{tabularx}
\usepackage{multirow}
\usepackage{supertabular}
\usepackage{url}
\graphicspath{{/stck/jgiehler/rans/postprocessing/figures/}{/stck/jgiehler/figures/}{/stck/jgiehler/experiments/postprocessing/figures/}}
\newcolumntype{x}[1]{>{\centering\arraybackslash\hspace{0pt}}p{#1}}
\DeclareSIUnit{\million}{\text{M.}}

\journal{Aerospace Science and Technology}

\begin{document}

\begin{frontmatter}

\title{Experimental and Numerical Investigation of Porous Bleed Control for Supersonic/Subsonic Flows, and Shock-Wave/Boundary-Layer Interactions}

\author{Julian Giehler}
\author{Thibault Leudiere}
\author{Robert Soares Morgadinho}
\author{Pierre Grenson}
\author{Reynald Bur}

\affiliation{organization={DAAA, ONERA, Institut Polytechnique de Paris},
            addressline={8 Rue des Vertugadins},
            city={Meudon},
            postcode={92190},
            country={France}}

\begin{abstract}

    This paper presents a comprehensive experimental and numerical investigation into the control of boundary layers using porous bleed systems. The study focuses on both supersonic and subsonic flow regimes, as well as on the control of shock-wave/boundary-layer interactions. By simplifying this complex problem into two separate scenarios, the research establishes the necessity of individually characterizing bleed systems for both supersonic and subsonic flow regimes due to variations in the flow topology within the bleed holes. In supersonic conditions, a strong agreement between experimental and numerical results validates the effectiveness of porous bleed in controlling boundary layers. Notably, three-dimensional effects are experimentally demonstrated, and variations of the boundary-layer profiles along the span are the first time experimentally proven. The study extends its scope to subsonic flows, revealing that while boundary-layer bleeding enhances flow momentum near the wall, mass removal induces a decrease in momentum in the outer boundary layer and external flow. The research also explores shock-wave/boundary-layer interaction control, achieving a remarkable alignment between simulations and experiments. The findings endorse the use of Reynolds-Averaged Navier-Stokes simulations for studying porous bleed systems in various flow conditions, providing valuable insights to enhance bleed models, especially in the context of shock-wave/boundary-layer interaction control.

\end{abstract}

\begin{keyword}
Flow Control \sep Porous Bleed \sep CFD \sep  Supersonic Flow  \sep Suction
\end{keyword}

\end{frontmatter}

\section{Nomenclature}

\noindent\textit{Latin symbols}

\begin{supertabular}{@{}l @{\quad=\quad} l@{}}
    $A$  & Area \\
    $A_o$  & (Circular) extraction area \\
    $A_{pl,ex}$  & Plenum exit throat area \\
    $D$  & Hole diameter \\
    $H$  & Boundary layer shape factor \\
    $L$  & Length \\
    $M$  & Mach number \\
    $\dot{m}$  & Mass flow rate \\
    $p$  & Static pressure \\
    $Q_{sonic,w}$  & Surface sonic flow coefficient \\
    $R$  & Specific gas constant (air)\\
    $Re$  & Reynolds number\\
    $T$  & Static temperature \\
    $T/D$  & Thickness-to-diameter ratio \\
    $TR$  & Throat ratio \\
    $v$  & Transpiration velocity \\
    $x,y,z$  & Cartesian coordinates in streamwise, wall-normal, and span wise direction \\
    $\hat{x}$, $\check{x}$  & Streamwise position from start/end of bleed region \\
\end{supertabular}\\

\noindent\textit{Greek symbols}

\begin{supertabular}{@{}l @{\quad=\quad} l@{}}
    $\alpha$  & Angle of attack \\
    $\beta$  & Stagger angle \\
    $\delta$ & Boundary layer thickness \\
    $\delta_1$  & Displacement thickness \\
    $\mu$  & Mach angle \\
    $\varepsilon_{\tau}$  & Rise in wall shear stress \\
    $\gamma$  & Heat capacity ratio \\
    $\phi$  & Porosity level \\
    $\rho$  & Density \\
    $\tau$  & Shear stress \\
\end{supertabular}\\

\noindent\textit{Subscripts}

\begin{supertabular}{@{}l @{\quad=\quad} l@{}}
    $bl$  & Bleed\\
    $c$  & Compressible\\
    $h$  & Hole\\
    $pl$  & Plenum\\
    $sonic$  & Sonic \\
    $t$  & Total \\
    $w$  & External wall\\
    $\infty$  & Free-stream\\
    $99$  & 99 \% boundary layer thickness \\
\end{supertabular}

\section{Introduction}

Porous bleed systems are a proven technology for controlling shock-wave/boundary-layer interactions~\cite{Babinsky2008}. The idea is to remove the low-momentum flow in the vicinity of the wall to make the boundary layer more resilient against adverse pressure gradients. Although the basic principle is simple, the flow physics inside the hole is difficult to understand and predict. The small size of the bleed holes, whose diameter is approximately the compressible boundary-layer displacement thickness~\cite{Syberg1973a, Harloff1996}, makes measurements inside the hole challenging if not impossible. In contrast, owing to the increase in computational power in previous years, numerical simulations investigating bleed flow have become feasible~\cite{Shih2008, Hamed2011, Oorebeek2015, Zhang2020b, Schwartz2023, Giehler2023b}.

However, numerical methods require validation and, therefore, a comprehensive database of experimental findings. Previous experimental studies were limited to a small range of geometrical parameters~\cite{Willis1995b, Willis1996} or a single-hole bleed~\cite{Eichorn2013}, which results in an overestimation of the bleed efficiency as the interaction between the holes is not considered~\cite{Giehler2023b}. Moreover, these experiments were limited to supersonic flows, whereas the flow topology was completely different for subsonic flows, leading to higher bleed mass fluxes~\cite{Harloff1996, Giehler2023a}. In contrast, experiments on a complex flow case with a shock-wave/boundary-layer interaction~\cite{Willis1995a} were performed without gaining detailed knowledge of the bleed behavior under subsonic conditions donwstream of a shock.

Another limitation of existing experiments is the assumption of a two-dimensional flow. Although simulations can be conducted under perfectly symmetric conditions, side wall effects and corner flows cannot be neglected in experiments. Also, the application of porous bleeds near a corner is a classical arrangement in supersonic air intakes, where often internal flows are controlled. Previous studies have shown that the use of flow control in the center of the working section leads to an increase in the size of the corner flows~\cite{Titchener2011a, Titchener2013}. Thus, even the flow at the center of the wind tunnel does not necessarily correspond to a two-dimensional flow. Moreover, the three-dimensionality of holes leads to variations in the flow field along the span wise direction, as shown by~\citet{Oorebeek2015} and~\citet{Giehler2023b}. In previous experiments, boundary-layer profiles were measured only at the center of the wind tunnel without considering these effects~\cite{Willis1996}.

In this study, we used a combination of experiments and three-dimensional steady-state Reynolds-averaged Navier-Stokes (RANS) simulations to address existing deficiencies. The experimental study of two porous plates targeted the validation of our numerical setup, which was used to generate a comprehensive database showing the strong effects of different geometrical parameters~\cite{Giehler2023b}. Moreover, three-dimensional effects, such as the impact of the center bleed on the corner flow and the flow field variation along the span~\cite{Hamed2011, Oorebeek2015, Giehler2023b}, are observed. Because the direct investigation of shock-wave/boundary-layer interactions is very complex, supersonic and subsonic boundary-layer bleeding on a flat plate are investigated separately before focusing on the control of an incident shock.

The paper is organized as follows: In Sec.~\ref{sec:method}, we introduce the experimental and numerical methodology. In the following section, the results are discussed, starting with boundary-layer control for supersonic and subsonic conditions before heading to the shock-wave/boundary-layer interaction. Finally, our conclusions are presented in the final section.

\section{Methodology} \label{sec:method}

In this section, both the experimental and numerical techniques are presented. The first investigated problem was the boundary-layer bleeding on a flat plate for super- and subsonic turbulent flows to determine the working principles of a porous bleed in both flow regimes. In the second step, a shock generator was included in the setup to investigate the shock-wave/boundary-layer interaction control. The domains were identical in the experimental and numerical approaches.

\begin{figure*}[t!]
    \centering
    \includegraphics[trim={0mm 2mm 0mm 2mm}, clip]{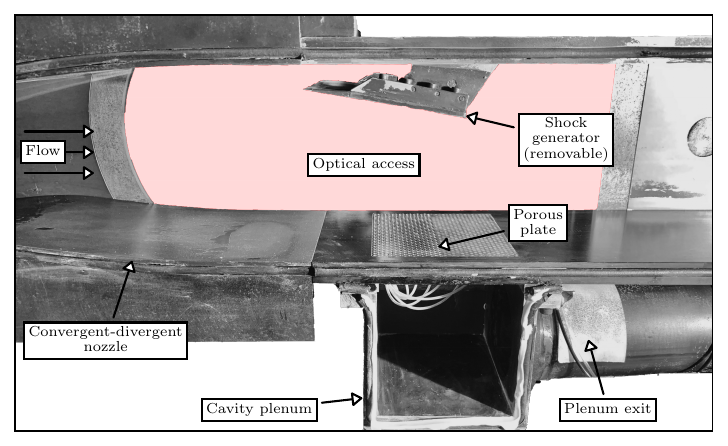}
    \caption{Experimental setup}
    \label{fig:experiment}
\end{figure*}

\subsection{Experimental setup}

\subsubsection{Wind tunnel}

The experiments were performed in the continuous-running S8Ch supersonic wind tunnel of the ONERA in Meudon, supplied with dried atmospheric air (see the total conditions in Tab.~\ref{tab:totalconditions}). Fig.~\ref{fig:experiment} shows a photograph of the working section. The convergent-divergent nozzle was designed to generate a Mach $M=1.62$ flow. The size of the working section was \SI{120}{\mm} in width and \SI{120}{\mm} in height. The porous bleed system was installed \SI{50}{\mm} downstream from the nozzle end. Windows on the sidewalls allow optical access to the entire region of interest around the bleed region.

In addition, a shock generator can be mounted in the working section. Its position and angle of attack are adjustable so that the incident shock wave is located on the porous bleed system.
A vent upstream of the compressor allowed the suction of air from a bypass channel. Thus, the mass flow rate passing the working section, and in turn, the Mach number can be varied for subsonic conditions, which was used in this study to also perform measurements for a fully subsonic flow inside the wind tunnel. A Mach number of $M=0.5$ was selected to avoid the appearance of a normal shock inside the divergent part of the nozzle due to choking of the nozzle throat and an under-expanded working regime.

\begin{table}[h!]
    \setlength{\tabcolsep}{4pt}
    \centering
    \begin{tabularx}{.5\linewidth}{x{0.15\textwidth}x{0.15\textwidth}x{0.12\textwidth}}
         \toprule
         & $p_t$ & $T_t$\\
         \midrule
         \multirow{3}{*}{Experiments} &\SI{97370}{\Pa} &\SI{290.6}{\K} \\
          & - & - \\
          & \SI{101960}{\Pa} &\SI{317.6}{\K} \\
         \midrule
         Simulations &\SI{93000}{\Pa} &\SI{300}{\K} \\
         \bottomrule
    \end{tabularx}
    \caption{Total conditions in experiments (range during all measurements) and simulations}
    \label{tab:totalconditions}
\end{table}

\subsubsection{Porous bleed system}

The porous plate had a length of \SI{40}{\mm} and a span of \SI{80}{\mm}, corresponding to 2/3 of the channel span. Preliminary numerical investigations have shown that wider plates do not affect the center flow but require higher bleed rates as the suction area is increased. Two porous plates are utilized in this study with hole diameters equal to $D=$ \SIlist{0.5;2.0}{\mm}. The plate with the smaller holes had a porosity of $\phi=$ \SI{14.5}{\%}, whereas the second plate had a porosity of $\phi=$ \SI{29.6}{\%}. Independent of the hole size, the thickness-to-diameter ratio was kept constant at $T/D=1$. The holes are distributed in a triangular shape, resulting in different streamwise positions for every second column of the holes. The stagger angle between the columns was $\beta=$ \SI{30}{\degree} for both the plates. Tab.~\ref{tab:plates} summarizes the geometric parameters of the porous plates.

\begin{table}[h!]
    \setlength{\tabcolsep}{4pt}
    \centering
    \begin{tabularx}{.5\linewidth}{x{0.08\textwidth}x{0.08\textwidth}x{0.08\textwidth}x{0.08\textwidth}x{0.08\textwidth}}
         \toprule
         Plate & $D$ [mm] & $\phi$ [\%] & $\beta$ [$^\circ$] & $T/D$ [-]\\
         \midrule
         AR & \num{0.5} & \num{14.5} & \num{30} & \num{1} \\
         HR & \num{2.0} & \num{29.6} & \num{30} & \num{1} \\
         \bottomrule
    \end{tabularx}
    \caption{Investigated porous plates}
    \label{tab:plates}
\end{table}

The bleed mass flow rate can be varied by using a choked throat at the plenum exit. The throat diameter varies from $D_{pl,ex}=$ \SIrange{10}{50}{\mm}, while the inner pipe diameter is \SI{76.2}{\mm} (\SI{3}{inches}). A stiffening system was installed inside the cavity to avoid bending of the plate caused by high-pressure differences.

\subsubsection{Laser-Doppler-Velocimetry}

A two-component Laser-Doppler-Velocimetry system was used to describe the flow field around the bleed region. Two lasers from the \textit{Coherent Genesis MX SLM-series} with wavelengths \SI{514.5}{\nm} and \SI{488}{\nm} were utilized to measure both streamwise and wall-normal velocity components of the flow. Both fringe patterns were tilted by \SI{45}{\degree} with respect to the flow to obtain the closest possible measurements to the wall. A \SI{40}{\mega\Hz} frequency shift is added to one of the beams using the \textit{FiberFlow} system from \textit{Dantec} to resolve negative velocities.

Both the emitted beams had a waist diameter of \SI{4.3}{\mm}. The distance between the beams was set to \SI{35}{\mm}, a beam expander with a ratio of \num{1.95} and a converging lens with a focal length of \SI{230}{\mm} were used. The probe volume characteristics can be calculated using classical relations~\cite{Durst1976}, resulting in probe diameters of \SI{35.04}{\um}/\SI{33.23}{\um}, probe lengths of \SI{0.46}{\mm}/\SI{0.44}{\mm}, and a fringe spacing of \SI{3.39}{\um}/\SI{3.22}{\um}. With the given characteristics, sampling rates on the order of $\mathcal{O}$(\SI{1}{\kHz}) were achieved.

\subsubsection{Flow visualization}

A Background Oriented Schlieren (BOS) system~\cite{Nicolas2016} was installed to monitor the flow in real time using a \SI{2.3}{MP} camera. Using the BOS method, the deviation of the light rays induced by density gradients was measured by cross-correlating one image of a random pattern without flow and one with the flow. The deflections in the streamwise and wall-normal directions were estimated using the in-house \textit{FOLKI} algorithm~\cite{Champagnat2011}.

\subsubsection{Pressure measurements}

Static pressure measurements were performed to monitor the flow. Pressure taps on the top wall inside the convergent-divergent nozzle and working section were used to check the correct starting of the working section. Moreover, pressure taps on the bottom wall and porous plate were installed with an offset of \SI{10}{\mm} from the center plane to acquire the wall pressure changes induced by boundary-layer bleeding. Inside the cavity plenum, pressure taps on the side and bottom walls were used to measure the plenum pressure, while pressure taps upstream and downstream of the sonic throat were used to prove choking conditions. With a hole diameter of $D=$ \SI{0.4}{\mm}, the size of the pressure taps is in the same order of magnitude as the bleed holes for plate AR.

\subsection{Numerical setup}

\subsubsection{Geometry and mesh}

\begin{figure}[h!]
    \centering
    \begin{subfigure}[t!]{.400\textwidth}
        \centering
        \includegraphics[width=\linewidth]{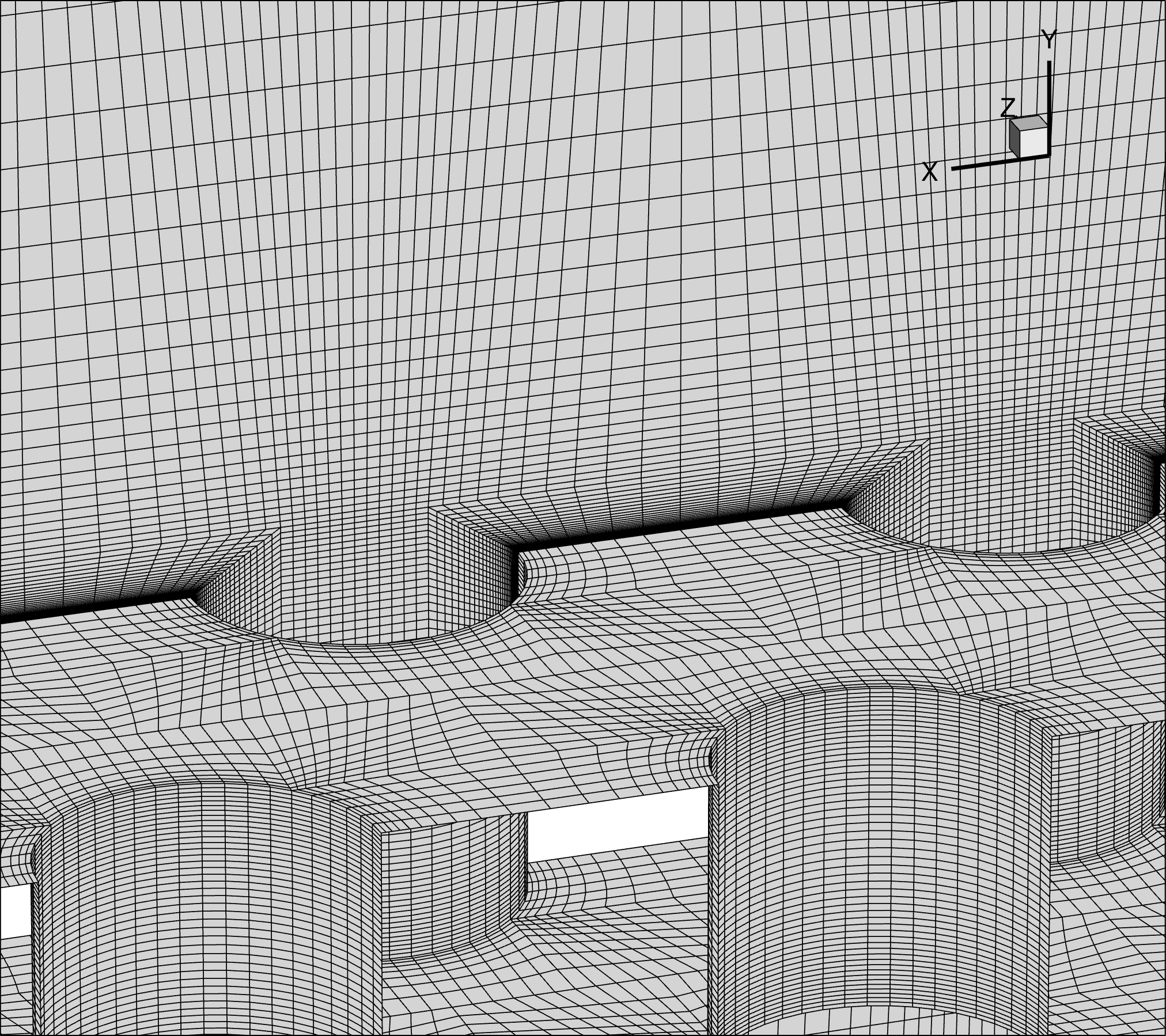}
        \caption{View on mesh around the holes}
        \label{fig:meshHole}
    \end{subfigure}
    \hspace{3mm}
    \begin{subfigure}[t!]{0.51\textwidth}
        \centering
        \includegraphics[width=\linewidth]{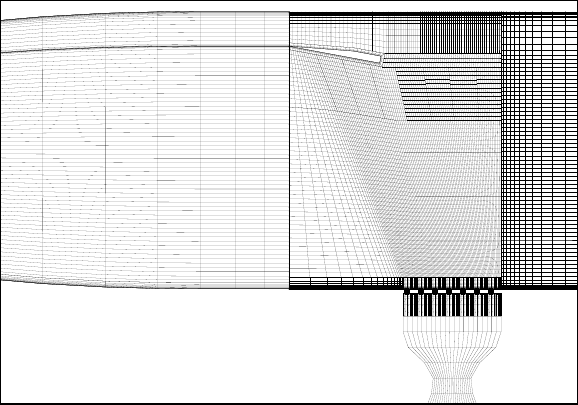}
        \caption{View on mesh with shock generator (every 4th grid line)}
        \label{fig:meshTunnel}
    \end{subfigure}
    \caption{Computational mesh}
    \label{fig:mesh}
\end{figure}

The entire working section was meshed using the in-house pre-processing tool and mesh-generator \textit{Cassiopee}~\cite{Benoit2015}. For the supersonic cases with and without the shock generator, the convergent-divergent nozzle was added to the domain. The mesh around the holes was fully parameterized, allowing the variation of several bleed parameters, such as the plate length, hole diameter, or plate thickness. Each hole was modeled out of five blocks using a butterfly mesh, as shown in Fig.~\ref{fig:mesh}. A C-grid topology around the hole walls was used to accurately resolve the boundary layer by simultaneously reducing the number of cells. The minimum wall-normal cell size was set as \SI{.2}{\um} ($y^+ \approx 1$) inside and outside the holes. A preliminary mesh sensitivity study was performed as part of our previous study~\cite{Giehler2023b} and showed that a cell-to-cell growth ratio of maximal \num{1.1} is required to accurately predict the local mass flow rate passing through the bleed holes.

The total number of cells was \SI{6.2}{\million} independent of the plenum pressure for plate AR and \SI{1.9}{\million} for plate HR in the quasi-two-dimensional (Q2D) setup used for the supersonic boundary-layer bleeding. Moreover, simulations of the wind tunnel half-span (WTHS) were conducted for plate HR and a span of the porous plate of \SI{40}{\mm}, resulting in \SI{55.9}{\million} cells. For the subsonic boundary-layer bleeding, the number of cells was reduced to \SI{1.2}{\million} and \SI{3.7}{\million}, respectively, because the convergent-divergent nozzle was not included in the mesh.
To observe the effect of an incident shock, a shock generator was included in the wind tunnel mesh using a C-grid topology. Thus, the number of cells increased to \SI{3.7}{\million} and \SI{10.7}{\million} cells, respectively. Again, simulations of the half-span were performed with a mesh size of \SI{106}{\million} cells.

\subsubsection{Flow solver}

The compressible Navier-Stokes equations were solved numerically using the ONERA-Safran finite-volume solver \textit{elsA}~\cite{Cambier2013}. The Spalart-Allmaras turbulence model with the quadratic constitutive relation~\cite{Spalart2000} was applied. The second-order-accurate Roe upwind scheme, combined with the min-mod limiter and the Harten entropic correction, was used as the spatial scheme, while a backward-Euler scheme was used for the time integration with local time-stepping. A subsonic velocity inlet condition was set upstream of the convergent-divergent nozzle, and the plenum exit was set as a supersonic outlet, along with the main outlet in the supersonic cases with and without shock-wave/boundary-layer interaction. For investigating the subsonic flow control, the main outlet was set as a pressure outlet, and the inlet was set to fit the experimentally acquired upstream boundary-layer profile. The top and bottom walls and walls around the holes were made no-slip walls, while a slip-wall was applied at the plenum side walls to reduce the mesh size inside the cavity. The domain was limited using a symmetry boundary condition on the front and back for the Q2D simulations, and for the WTHS simulations, the center plane was set as a symmetry boundary condition, while the back wall was treated as a no-slip wall.

Contrary to the experiments, the bleed mass flow rate was fixed by a choked nozzle instead of a choked throat. Therefore, the cavity plenum was prolonged, and its exit was shaped as a convergent-divergent nozzle to produce supersonic conditions at the outlet. The size of the chocked plenum exit area $A_{pl,ex}$ was determined with respect to the bleed area $A_{bl}$, which was the sum of all bleed hole areas $A_{bl,i}$, by using the throat ratio

\begin{equation}
    TR = \frac{A_{pl,ex}}{A_{bl}} = \frac{A_{pl,ex}}{\sum A_{bl,i}}.
\end{equation}
For the comparison with the experiments, the static plenum pressure $p_{pl}$ was extracted at a distance of three hole diameter to the exit of the holes inside the cavity.

\section{Results} \label{sec:results}

In the following, the experimental and numerical findings are compared. In the first step, the supersonic turbulent boundary-layer bleeding is investigated before heading to subsonic conditions, and in the final step to the control of the shock-wave/boundary-layer interaction.

\subsection{Supersonic boundary-layer bleeding}

\subsubsection{Flow field analysis} \label{sec:flowField}

\begin{figure*}[!hbt]
    \centering
    \begin{subfigure}[t!]{1.00\textwidth}
        \centering
        \makebox[\textwidth][c]{\includegraphics[trim={0mm 50mm 0mm 17mm}, clip]{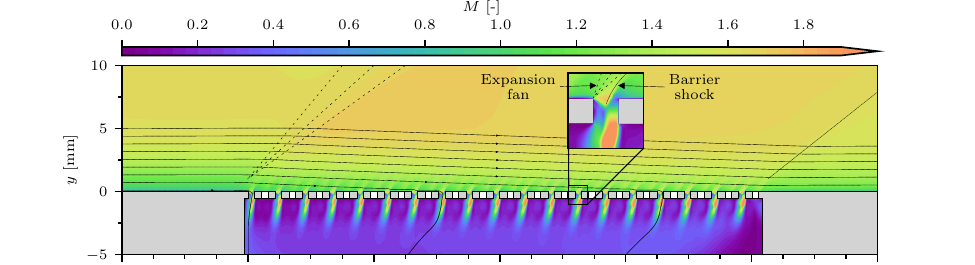}}
        \caption{Plate AR ($p_{pl}/p_t = 0.044$)}
        \label{fig:contour05}
    \end{subfigure}
    \begin{subfigure}[t!]{1.00\textwidth}
        \centering
        \makebox[\textwidth][c]{\includegraphics[trim={0mm 2mm 0mm 66mm}, clip]{ASTtopologySupersonic.pdf}}
        \caption{Plate HR ($p_{pl}/p_t = 0.076$)}
        \label{fig:contour20}
    \end{subfigure}
    \caption{Mach contour fields; white areas illustrate holes of the second column}
    \label{fig:contour}
\end{figure*}

The flow topology in the vicinity and inside the holes is shown for the Q2D simulations with both porous plates in Fig.~\ref{fig:contour}. The chosen working regime is characterized by choked holes, caused by a low pressure ratio from external wall to cavity plenum. Thus, the bleed operates close to the limiting conditions, i.e., the maximum bleed rate. In Fig.~\ref{fig:contour05}, the flow field for plate AR with $D=$ \SI{0.5}{\mm} is illustrated. Following the streamlines, a deflection of the flow towards the wall is noted at the beginning of the plate as part of the boundary layer is sucked. This leads to an increase in the Mach number as the flow is accelerated. The so-called trailing shock is located at the end of the plate, where the porous plate ends, redirecting the flow in the wall-parallel direction.

Close to the wall, further expansion and shock waves are generated by the flow streaming into the holes, as seen in the zoom view in Fig.~\ref{fig:contour}. At each hole front, an expansion fan is located, bending the flow inside the hole. Further downstream, the barrier shock returns the flow parallel to the wall. Inside the hole, the flow is choked, leading to a supersonic under-expanded jet into the cavity plenum.

A similar flow field is apparent for plate HR, as shown in Fig.~\ref{fig:contour20}. In comparison to the first plate, the thinning of the boundary layer is more prominent as highlighted by the stronger bending of the streamlines. This is a consequence of the higher porosity level, which is approximately twice the one for plate AR, and in turn, the higher bleed mass flow rate as the open area is increased. Moreover, the local flow phenomena induced by the bleed holes have a more significant effect on the flow field and penetrate deeper inside the boundary layer. The higher momentum of the captured flow is the reason since a higher flow momentum results in stronger barrier shocks~\cite{Giehler2023b}.

\begin{figure*}[!hbt]
    \centering
    \begin{subfigure}[t!]{1.00\textwidth}
        \centering
        \makebox[\textwidth][c]{\includegraphics[trim={0mm 44mm 0mm 2mm}, clip]{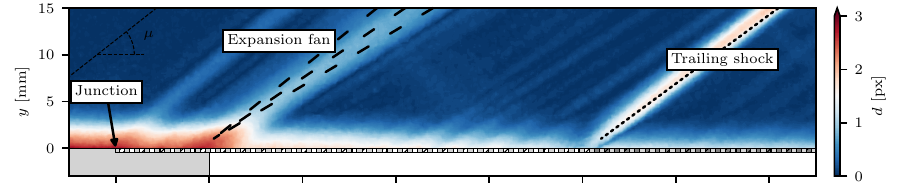}}
        \caption{Plate AR ($p_{pl}/p_t = 0.046$)}
        \label{fig:bos05}
    \end{subfigure}
    \begin{subfigure}[t!]{1.00\textwidth}
        \centering
        \makebox[\textwidth][c]{\includegraphics[trim={0mm 2mm 0mm 36mm}, clip]{ASTbosSupersonic.pdf}}
        \caption{Plate HR ($p_{pl}/p_t = 0.073$)}
        \label{fig:bos20}
    \end{subfigure}
    \caption{BOS visualization of the flow around the porous plate}
    \label{fig:bos}
\end{figure*}

The flow fields observed for both porous plates using the BOS method are illustrated in Fig.~\ref{fig:bos}. Both the expansion fan at the beginning of the plate and the trailing shock at the end are visible. Moreover, as the displacement $d$ is the highest close to the wall where the density gradient is the maximum, the thinning of the boundary layer is perceptible. Downstream of the bleed region, the boundary layer is significantly fuller and smaller in both cases.

On the porous plate, the effect of the hole flow is apparent. Especially for plate HR with $D=$ \SI{2.0}{\mm} holes, the effect on the flow inside the boundary layer is observed as expansion and compression waves are evident. Since the porous plate is mounted on the floor, an overlapping of the pocket and the plate closes the holes located at the first \SI{10}{\mm} of the plate. Thus, no suction is present in this region, but the hole contour creates roughness. Also, a Mach wave caused by the junction from the floor to the porous plate is present. The holes downstream of the bleed region are closed with plaster, which induces further Mach waves that do not affect the flow so that the measurement of the boundary-layer profiles in this region is feasible. The appearance of these Mach waves is more prominent for the larger holes as the filled area is increased.

Altogether, the same flow topology is found in both experiments (Fig.~\ref{fig:bos}) and simulations (Fig.~\ref{fig:contour}). However, expansion fans and shock waves are more smeared using the BOS visualization compared to the simulations. Contrary to a Schlieren visualization, no mirrors are used to obtain parallel light rays passing the wind tunnel. Thus, the view is only aligned with the span wise coordinate in the center of the image. Consequently, the deflection is higher in the center, resulting in occasional bad correlations, as visible in Fig.~\ref{fig:bos20}.

\begin{figure*}[!hbt]
    \centering
    \begin{subfigure}[t!]{1.00\textwidth}
        \centering
        \makebox[\textwidth][c]{\includegraphics[trim={0mm 89mm 0mm 2mm}, clip]{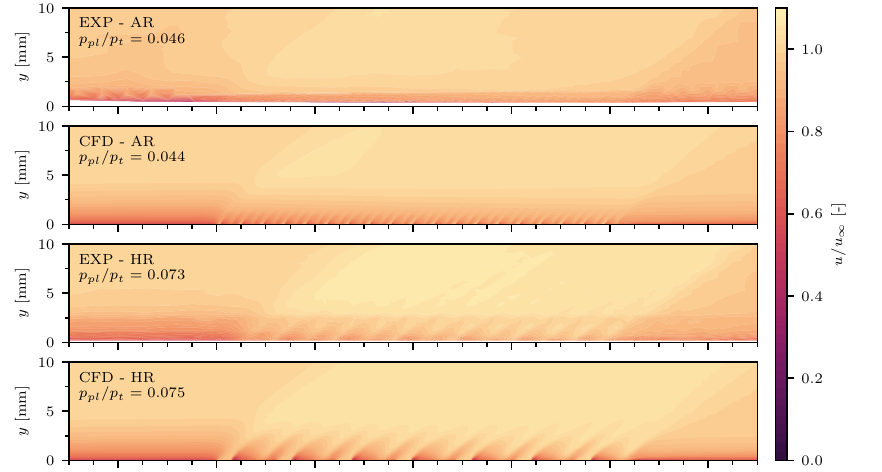}}
        \caption{Streamwise velocity component $u$}
        \label{fig:ldvcfdu}
    \end{subfigure}
    \begin{subfigure}[t!]{1.00\textwidth}
        \centering
        \makebox[\textwidth][c]{\includegraphics[trim={0mm 2mm 0mm 82mm}, clip]{ASTflowFieldSuper.pdf}}
        \caption{Wall-normal velocity component $v$}
        \label{fig:ldvcfdv}
    \end{subfigure}
    \caption{Comparison of flow field from LDV measurements and simulations}
    \label{fig:ldvcfd}
\end{figure*}

For a more quantitative comparison, the whole flow field measured by means of LDV is shown against the numerical results in Fig.~\ref{fig:ldvcfd} in streamwise (Fig.~\ref{fig:ldvcfdu}) and wall-normal direction (Fig.~\ref{fig:ldvcfdv}). Both velocity components are normalized by the free-stream velocity to eliminate temperature variations that occurred between the different runs in the experiments.

A look at Fig.~\ref{fig:ldvcfdu} illustrates a good fit of the streamwise component of the flow for both hole diameters. The effect of the bleed on the boundary layer is well modeled in the simulations for both plates. In the case of the small holes, the LDV measurements are not able to fully resolve the local flow phenomena around the holes, which is caused by the size of the measurement volume in the span wise direction of approximately one hole diameter. On the contrary, the flow field is in the simulations directly extracted on the symmetry plane. Moreover, a Mach wave induced by the junction of the plates at $x=$ \SI{40}{mm} is apparent in the experimental data, leading to a slight thickening of the boundary layer. In contrast, no Mach wave is found in the LDV data for the larger holes, where the inflow is more homogeneous compared to the measurements for the $D=$ \SI{0.5}{\mm} plate. Moreover, the expansion fans and barrier shocks induced by the bleed holes are resolved since the hole diameter is in this case significantly larger than the probe volume, and the phenomena are more pronounced.

Also, the wall-normal velocity component is well reproduced by the RANS simulations, as shown in Fig.~\ref{fig:ldvcfdv}. A relatively low deflection of the flow towards the wall is found for the small holes compared to the large holes, caused by the lower porosity level. Moreover, the flow field is more homogeneous for the small hole diameters. In contrast, penetration of the expansion fans and barrier shocks even on the flow outside the boundary layer is notable. This effect is visible in the experimental data for the $D=$ \SI{2.0}{\mm} holes. In comparison, the flow field acquired from the simulations looks more homogeneous.

\begin{figure*}[!bht]
    \centering
    \makebox[\textwidth][c]{\includegraphics[trim={0mm 2mm 0mm 2mm}, clip]{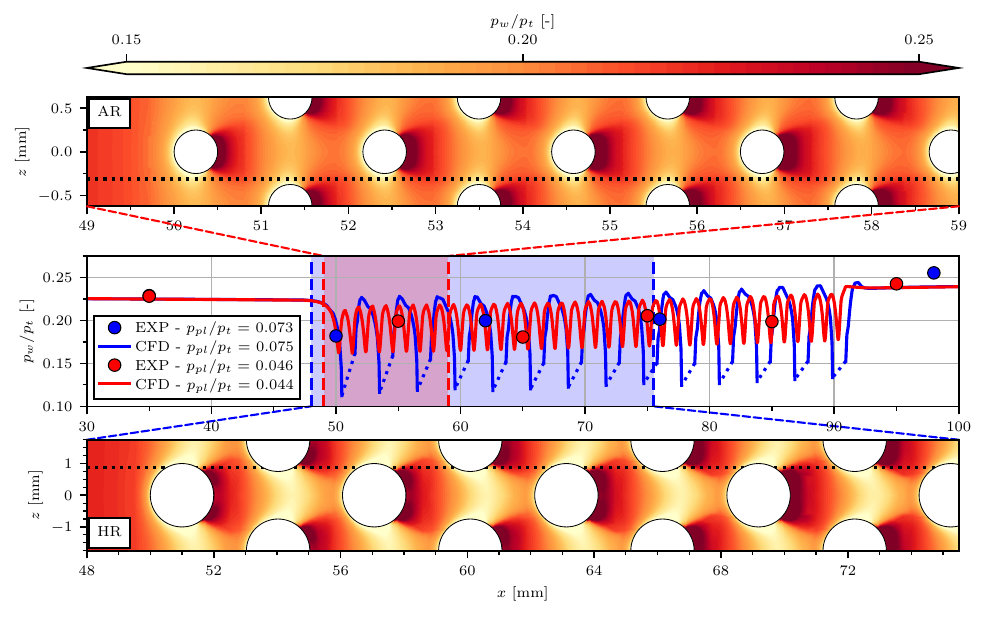}}
    \caption{Static wall pressure along the bleed region}
    \label{fig:pressure}
\end{figure*}

The wall pressure along the bleed region is presented in Fig.~\ref{fig:pressure}. On the top, the numerically extracted pressure contour on the wall is illustrated for the first bleed holes with a diameter of $D=$ \SI{0.5}{\mm}. The suction effect is evident: near the front of the holes, the pressure decreases because of the presence of an expansion fan. Further downstream, the so-called barrier shock inside the hole leads to an adverse pressure gradient resulting in a significantly higher wall pressure downstream of the hole. On the bottom, the porous plate HR is shown, illustrating the same phenomena. However, the wall pressure below the holes is significantly lower, which is a result of the lower relative distance between the holes, and the consequently more effective flow control~\cite{Giehler2023b}.

For a comparison with the experiment, the wall pressure is extracted on a line between two hole columns (dotted line), where the wall is fully solid for the plate with $D=$ \SI{0.5}{\mm} holes, and largely fully for the second plate. The plot in the center details the trend of the wall pressure along these lines. At the beginning of the plate, where an expansion fan is located, the wall pressure drops. Along the porous plate, the pressure fluctuates caused by the flow phenomena induced by the bleed holes. These fluctuations are more prominent for the plate with the large holes. However, positive slopes along the plates are notable due to the compression along the porous plates. Since the boundary-layer thinning is not linear but stronger at the beginning of the plate, the flow is continuously deflected towards the wall-normal direction~\cite{Giehler2023b}. At the end of the bleed region, the pressure increases because of the trailing shock, resulting in a slightly increased wall pressure downstream of the plate compared to the upstream pressure.

In the experiments, the wall pressure is measured at discrete locations around and in the bleed region, as illustrated by the points. Upstream and downstream of the bleed region, the measured wall pressure is slightly higher, which may be caused by the three-dimensionality of the flow inside the wind tunnel or Mach waves caused by the junction between the working section and the convergent-divergent nozzle. However, the deviation equals upstream and downstream, which proves the same behavior of the bleed system. Along the plate, the measured pressure is within the range of the pressure fluctuations. The non-steady trend of the experiments results from the diameter of the pressure taps and their positions with respect to the bleed holes. As the diameter is of the same order of magnitude as the smaller bleed hole diameter, the pressure is averaged above a relatively large area. Depending on the position, the acquired pressure is mainly affected by the expansion fan or barrier shock and thus lower or higher.

Overall, the numerically and experimentally observed flow fields are similar. The same flow topology is apparent, and the flow is not strongly affected by any geometrical differences.


\subsubsection{Boundary layer profiles at the center}

\begin{figure*}[!bht]
    \centering
    \makebox[\textwidth][c]{\includegraphics[trim={0mm 2mm 0mm 2mm}, clip]{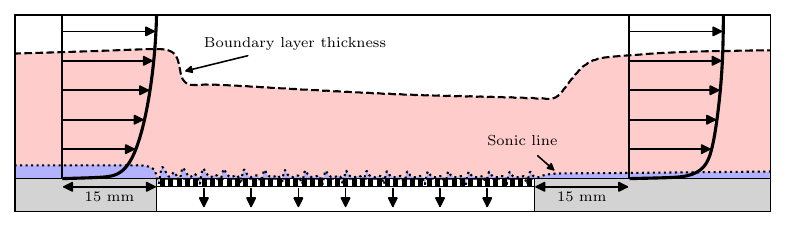}}
    \caption{Illustration on the effect of the porous bleed system on the boundary layer based on numerical results}
    \label{fig:thinning}
\end{figure*}

In the following, the boundary-layer profiles are compared. Fig.~\ref{fig:thinning} schematically visualizes the location of the observed profiles. The profiles are measured \SI{15}{\mm} upstream and downstream of the plate in both simulations and experiments. Upstream of the plate, this distance is chosen as the position is unaffected by the junction from the floor to the porous plate and the non-suction holes. Moreover, the upstream influence of the porous bleed is numerically found to be below \SI{5}{\mm}, which results in similar boundary-layer profiles independent of the bleed rate.

As described in Sec.~\ref{sec:flowField}, a trailing shock is caused by a deflection of the flow at the end of the porous bleed. Consequently, the boundary layer downstream of the bleed region is disturbed by the shock and does not allow accurate measurements. Therefore, the same distance of \SI{15}{\mm} is selected here to assess the effect of bleeding on the boundary-layer profile. A view in Fig.~\ref{fig:thinning} reveals that the boundary-layer thickness is slightly lower downstream of the trailing shock. However, the sonic height is significantly lower, resulting in a fuller boundary layer.

\begin{figure}[!thb]
    \centering
    \makebox[\textwidth][c]{\includegraphics[trim={0mm 2mm 0mm 2mm}, clip]{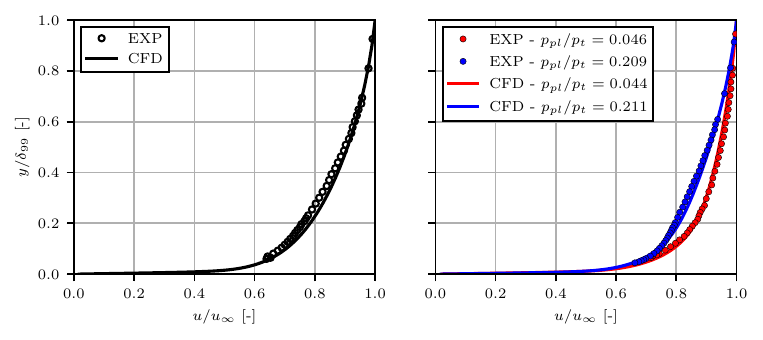}}
    \begin{subfigure}[t!]{.40\textwidth}
        \centering
        \caption{\SI{15}{\mm} upstream of the bleed region}
        \label{fig:inflow}
    \end{subfigure}
    \begin{subfigure}[t!]{.40\textwidth}
        \centering
        \caption{\SI{15}{\mm} downstream of the bleed region}
        \label{fig:downflow}
    \end{subfigure}
    \caption{Comparison of boundary-layer profiles for plate AR from experiments and simulations}
    \label{fig:profilesValidation}
\end{figure}

Fig.~\ref{fig:profilesValidation} compares the velocity profile extracted from the simulation and the measured profile using LDV. The closest point to the wall experimentally measured is located at a wall-normal distance $y=$ \SI{0.2}{\mm} ($y^+=\mathcal{O}(100)$). The simulation fits very well with the experimental measurements. The boundary-layer thickness $\delta_{99}=$ \SI{4.3}{\mm} is equal in both experiments and simulations, even though the total conditions are slightly different (see Tab.~\ref{tab:totalconditions}).

The boundary-layer profile upstream of the bleed region (Fig.~\ref{fig:inflow}) extracted from the simulations is slightly fuller than the extracted profile from the experiments. This is probably caused by the higher wall roughness in the experimental setup. However, the differences are negligibly minor.

With the aim of validating the simulations for different working conditions of the porous bleed, the boundary-layer profiles downstream of the bleed region are measured for several bleed rates. Comparison is made with the simulation for the same pressure in the cavity plenum. As the flow at the center plane is assumed to be two-dimensional, the ratio of plenum pressure to total pressure is more suitable than using the bleed mass flow rate, which may be affected by the three-dimensionality of the flow inside the wind tunnel close to the side walls. The pressure is expected to be more uniform than the bleed mass flux.

The comparison of the simulations and the experiments downstream of the bleed region is detailed in Fig.~\ref{fig:downflow} for the lowest and the highest bleed rates. For the lowest bleed rate, the plenum pressure is slightly lower than the static wall pressure inside the channel \mbox{($p_w/p_t \approx 0.226$)}. Both experiments and simulations show the same trend again. However, the simulations predict a slightly fuller profile which is an artifact of the fuller inflow profile. Remarkably, the boundary-layer thickness $\delta_{99} =$ \SI{4.8}{\mm} is thicker than upstream of the bleed region, which is caused by the trailing shock.

For the highest bleed rate, the holes are choked, as shown by the simulations in Fig.~\ref{fig:contour}. The boundary-layer profiles obtained from simulations and experiments show an almost perfect fit in this case. The boundary layer is significantly fuller than in the case of a low bleed rate. Also, the boundary layer is thinner than the upstream profile with $\delta_{99}=$ \SI{3.7}{\mm}.

\begin{figure*}[!tbh]
    \centering
    \includegraphics[trim={0mm 2mm 0mm 2mm}, clip]{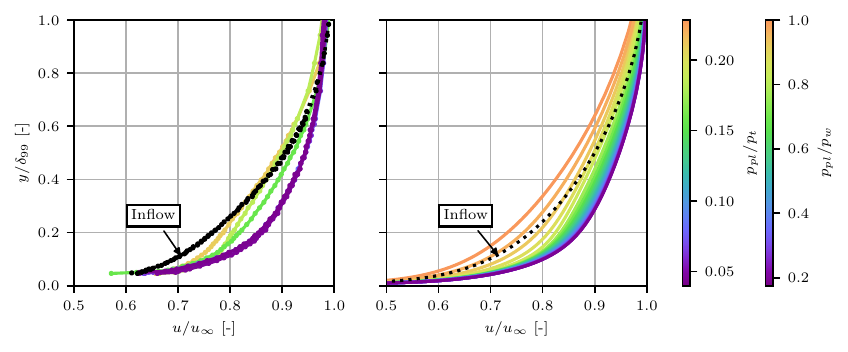}
    \begin{subfigure}[t!]{0.32\textwidth}
        \centering
        \caption{Experiments}
        \label{fig:profilesDownExp}
    \end{subfigure}
    \begin{subfigure}[t!]{0.40\textwidth}
        \centering
        \caption{Simulations}
        \label{fig:profilesDownCFD}
    \end{subfigure}
    \caption{Boundary layer profiles \SI{15}{\mm} downstream of the bleed region for plate AR}
    \label{fig:profilesDown}
\end{figure*}

All experimentally acquired boundary-layer profiles are compared in Fig.~\ref{fig:profilesDownExp} for the $D=$ \SI{0.5}{\mm} plate. Wall-normal coordinate and streamwise velocity are normalized with the inflow boundary-layer thickness and free-stream velocity. The significant influence of the bleed rate is apparent: The higher the bleed rate, the lower the pressure ratio and the fuller the boundary-layer profile. A significantly fuller boundary layer is already found for the lowest bleed rate. However, the suction mainly affects the lower part of the boundary layer, while the velocity is lower with further wall distance compared to the inflow. 

A further increase in the bleed rate leads to an even fuller boundary-layer profile. Again, the major difference is found in the near-wall region, while the outer boundary layer is mainly unaffected. For more significant bleed rates, the whole boundary layer is fuller. From a pressure ratio $p_{pl}/p_t \approx 0.1$, the profiles converge, which means that the maximum effect is achieved, and a lower pressure inside the cavity does not lead to fuller profiles. This results from choking the flow inside the holes, which limits the bleed mass flow rate.

The same trends are found in the numerical simulations, as shown in Fig.~\ref{fig:profilesDownCFD}. Low bleed rates mainly affect the near-wall region, while the outer boundary layer is unaffected, and natural boundary-layer growth is found. Lower pressure ratios increase the effectiveness of the porous bleed system until choking is achieved, which limits the bleed effect. Moreover, a certain amount of bleed is required to sustain the inflow conditions and to prevent natural boundary-layer growth, which is additionally increased by the induced roughness due to the porous plate geometry.

\begin{figure*}[!tbh]
    \centering
    \includegraphics[trim={0mm 2mm 0mm 2mm}, clip]{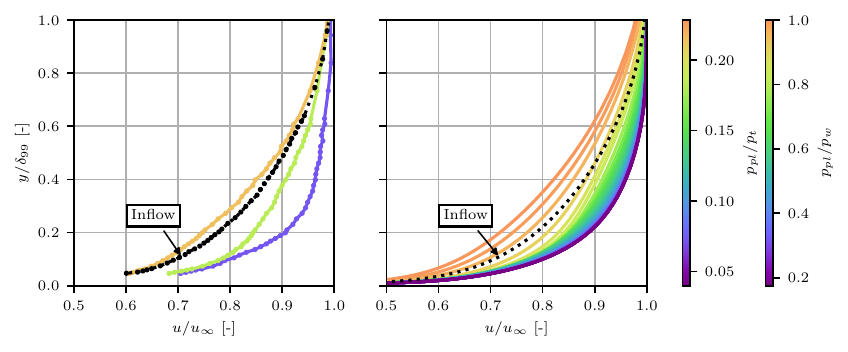}
    \begin{subfigure}[t!]{0.32\textwidth}
        \centering
        \caption{Experiments}
        \label{fig:profilesDownExpHR}
    \end{subfigure}
    \begin{subfigure}[t!]{0.40\textwidth}
        \centering
        \caption{Simulations}
        \label{fig:profilesDownCFDHR}
    \end{subfigure}
    \caption{Boundary layer profiles \SI{15}{\mm} downstream of the bleed region for plate HR}
    \label{fig:profilesDownHR}
\end{figure*}

The same comparison is found for the plate with the $D=$ \SI{2.0}{\mm} holes in Fig.~\ref{fig:profilesDownHR}. On the left side (Fig.~\ref{fig:profilesDownExpHR}), the experimentally acquired profiles are shown for three different bleed rates. In the case of the lowest bleed rate, corresponding to the highest pressure ratio, a degradation of the boundary layer with respect to the inflow is found. Thus, the bleed mass flow rate is not enough to prevent boundary-layer growth. For higher mass removals, the boundary-layer profiles are significantly fuller. Again, the positive effect on the boundary layer first appears in the near-wall region and is found for higher mass flow rates also with further distance to the wall.

RANS simulations of the same plate validate the experimental findings, as shown in Fig.~\ref{fig:profilesDownCFDHR}. In comparison to the first plate, we see a more significant reduction in the boundary-layer thickness, which is a result of the higher effectiveness caused by the higher porosity level~\cite{Giehler2023b}.


\subsubsection{Three-dimensionality of the flow} \label{sec:sup3d}

\begin{figure}[!tbh]
    \centering
    \begin{subfigure}[t!]{1.00\textwidth}
        \centering
        \makebox[\textwidth][c]{\includegraphics[trim={0mm 93mm 0mm 22mm}, clip]{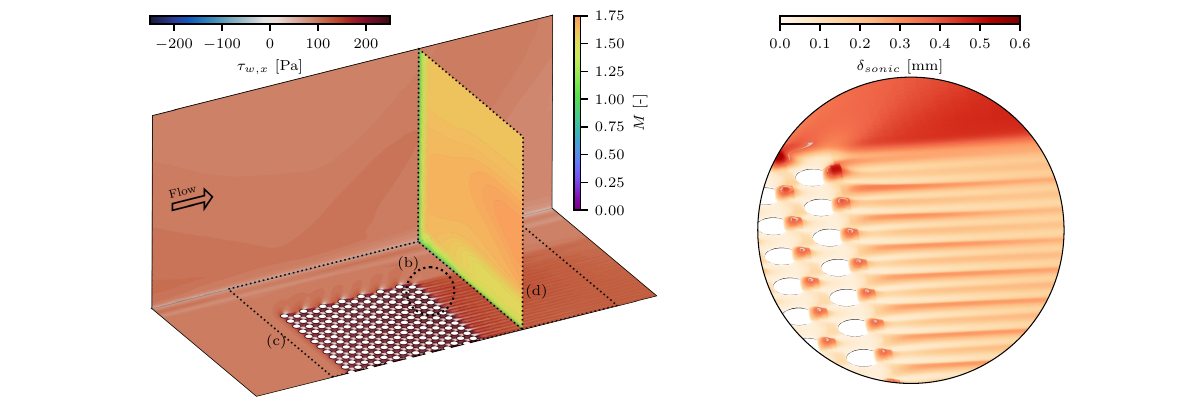}}
    \end{subfigure}
    \begin{subfigure}[t!]{0.64\textwidth}
        \centering
        \caption{Flow field inside the wind tunnel}
        \label{fig:windTunnela}
    \end{subfigure}
    \begin{subfigure}[t!]{0.35\textwidth}
        \centering
        \caption{Sonic height downstream of the bleed region}
        \label{fig:windTunnelb}
    \end{subfigure}
    \begin{subfigure}[t!]{1.00\textwidth}
        \centering
        \makebox[\textwidth][c]{\includegraphics[trim={0mm 12mm 0mm 104mm}, clip]{ASTflowWT.pdf}}
    \end{subfigure}
    \begin{subfigure}[t!]{0.64\textwidth}
        \centering
        \caption{Wall shear stress around the bleed region}
        \label{fig:windTunnelc}
    \end{subfigure}
    \begin{subfigure}[t!]{0.35\textwidth}
        \centering
        \caption{Flow field downstream of the bleed region}
        \label{fig:windTunneld}
    \end{subfigure}
    \caption{Flow topology inside the wind tunnel for supersonic boundary-layer bleeding from simulation ($p_{pl}/p_t=0.073$)}
    \label{fig:windTunnel}
\end{figure}

The three-dimensionality of the flow field has been already proven in previous numerical~\cite{Hamed2011, Wukie2015, Giehler2023b, Giehler2023c} and experimental~\cite{Oorebeek2015} studies. However, experiments showing the impact of the hole geometry on the boundary layer along the span are missing. In our previous numerical study~\cite{Giehler2023b}, we could show a significant variation of the flow field along the span even in a Q2D flow field. Particularly large hole diameters lead to inhomogeneous flow fields since the inter-hole distance is large, and a high-momentum captured flow results in strong compressible effects. For the investigation of three-dimensional effects, we focus on plate HR with $D=$ \SI{2.0}{\mm} holes as the flow field for small holes is too homogeneous and in the range of the experimental error. Moreover, the smaller amount of bleed holes in the case of large diameters makes simulations of the WTHS feasible in terms of the required mesh size.

Fig.~\ref{fig:windTunnel} illustrates the flow inside the wind tunnel off the center plane. Effects of the corner flow, but also variations of the flow field along the span downstream of the bleed region are significant. A general overview of the flow field is given in Fig.~\ref{fig:windTunnela}, visualizing the wall shear stress on the bottom and side wall of the wind tunnel, as well as a slice showing the flow field at a cross-section downstream of the bleed region, corresponding to the distance where the velocity profiles are extracted. Since the porous bleed covers only two-thirds of the span, the seemingly low impact on the corner flow is visible. Moreover, downstream of the bleed region, streaks of higher and lower wall shear stress are present.

A closer look at the last row of holes and the region downstream can be seen in Fig.~\ref{fig:windTunnelb}. Around the holes, the sonic height is extracted as a characteristic measure of the boundary-layer health. With regard to (strong) shock-wave/boundary-layer interactions, the sonic height can be directly linked to the interaction length and is therefore of major importance~\cite{Babinsky2011}. As seen in the figure, the sonic height varies significantly along the span. Downstream of the holes, a drastic increase is noted as the barrier shock induces a strong adverse pressure gradient. Also, a second streak of large sonic heights between the holes is found. Moreover, a low span wise influence of the porous bleed on the corner flow apparent. The flow close to the side walls seems to be unaffected as the sonic height is not reduced.

A better view at the effect on the corner flow is shown in Fig.~\ref{fig:windTunnelc}. As already stated before, the boundary-layer suction generates an expansion as the flow is redirected towards the wall. However, also flow from the side is captured, leading to the propagation of expansion waves in the span wise direction. As a consequence, the flow is redirected from the side walls to the bleed, as highlighted by the streamlines. The iso-contours of a deflection angle of $\theta=$ \SI{1}{\degree} illustrate the location of the first expansion waves, but also the propagation of the trailing shock further downstream at the plate end.
Consequently, the application of the porous bleed on only two-thirds of the wind tunnel span leads to an increase in the size of the corner flow, which may result in a need for further flow control in these regions~\cite{Titchener2013}. Since the flow-momentum decreases inside the corner, the flow is more vulnerable to separate if adverse pressure gradients are present.

Moreover, the footprints of the varying boundary-layer profiles along the span are found in the wall shear stress. The same streaks as for the sonic height are found. Directly downstream of the holes, as well as between the holes, the wall shear stress is significantly lower. The pattern of the footprint of the wall shear stress fits also very well the pattern found in the oil-flow visualization of~\citet{Oorebeek2013} for a similar plate geometry, but a higher Mach number of $M=2.5$.

The cross-section of the wind tunnel is shown in Fig.~\ref{fig:windTunneld}. Here again, the propagation of the expansion fan in the span wise direction is notable, leading to an increase in the Mach number, as well as the propagation of the trailing shock. Moreover, the sonic line as well as the boundary-layer thickness, based on the \SI{99}{\%} of the free-stream velocity at the inflow, are plotted as characteristic measures to evaluate the boundary-layer health. The trend of the boundary-layer thickness reveals apparently a thickening of the boundary layer. However, this is caused by a decrease in the Mach number as a result of the losses due to several barrier shocks and the trailing shock. Consequently, the flow-momentum outside of the boundary layer is smaller. In contrast, the sonic height is reduced compared to the flow at the side walls, as visible in the zoom-view. As a result, the porous bleed achieved the aim of increasing the flow-momentum inside the boundary layer.

\begin{figure}[t]
    \centering
    \begin{subfigure}[t!]{1.00\textwidth}
        \centering
        \makebox[\textwidth][c]{\includegraphics[trim={0mm 32mm 0mm 2mm}, clip]{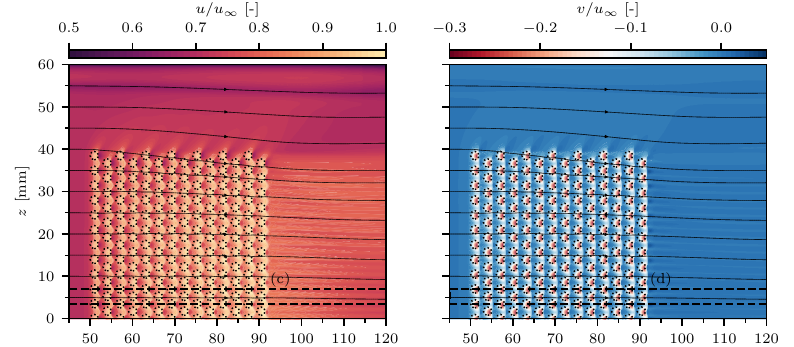}}
    \end{subfigure}
    \begin{subfigure}[t!]{0.45\textwidth}
        \centering
        \caption{Streamwise component along the span (CFD)}
        \label{fig:horizontala}
    \end{subfigure}
    \begin{subfigure}[t!]{0.45\textwidth}
        \centering
        \caption{Wall-normal component along the span (CFD)}
        \label{fig:horizontalb}
    \end{subfigure}
    \begin{subfigure}[t!]{1.00\textwidth}
        \centering
        \makebox[\textwidth][c]{\includegraphics[trim={0mm 17mm 0mm 61mm}, clip]{ASThorizontalPlane.pdf}}
        \caption{Streamwise component for simulations and experiments}
        \label{fig:horizontalc}
    \end{subfigure}
    \begin{subfigure}[t!]{1.00\textwidth}
        \centering
        \makebox[\textwidth][c]{\includegraphics[trim={0mm 2mm 0mm 77mm}, clip]{ASThorizontalPlane.pdf}}
        \caption{Wall-normal component for simulations and experiments}
        \label{fig:horizontald}
    \end{subfigure}
    \caption{Velocity contours at a wall-normal distance of $y/D=0.25$ ($y/\delta_{99}\approx 0.12$) for experiments ($p_{pl}/p_t=0.072$) and simulations ($p_{pl}/p_t=0.073$); plots in (c) and (d) show zoom-view to area highlights by dashed lines}
    \label{fig:horizontal}
\end{figure}

For the validation of the numerical findings, LDV measurements were performed at a horizontal plane with a distance of $y/D=0.25$. At the same height, the velocity components in streamwise and wall-normal directions are extracted. Fig.~\ref{fig:horizontal} shows both the numerical and experimental results. Since the measurements were acquired aside of the center plane, the velocity components are compared to the full simulation including the side walls. Again, a bending of the streamlines towards the wind tunnel center is found in Fig.~\ref{fig:horizontala} and~\ref{fig:horizontalb}, even downstream of the bleed region.

For the comparison with the experiments, we focus on the area between three columns of holes while the center column is staggered. The observed plane is acquired off the center plane. Fig.~\ref{fig:horizontalc} shows the comparison of simulations and experiments for the streamwise component, which reveal the same flow characteristics. The holes induce expansion waves, leading to an increase in the flow velocity before passing the barrier shocks. Downstream of the last holes, the streaks are noted, as well as their bending towards the wind tunnel center. In contrast to the simulations, the measured velocity differences are smaller, which is a result of the lower resolution of the measurements caused by the size of the measurement volume and the smaller amount of points. Consequently, the results are more smeared because of larger averaging areas. Please note, that the length of the porous plate differs slightly between experiments and simulations as the staggered column has one hole less in the experiments. However, previous findings~\cite{Giehler2023b} showed only marginal differences at these plate lengths.

Also, the measurements in the wall-normal direction fit well with the simulations (see Fig.~\ref{fig:horizontald}). As stated before, the streaks are less characteristic than for the streamwise component. Again, the experimentally measured variations are smaller than in the simulations because of the lower resolution.

\begin{figure*}[!tbh]
    \centering
    \includegraphics[trim={0mm 2mm 0mm 2mm}, clip]{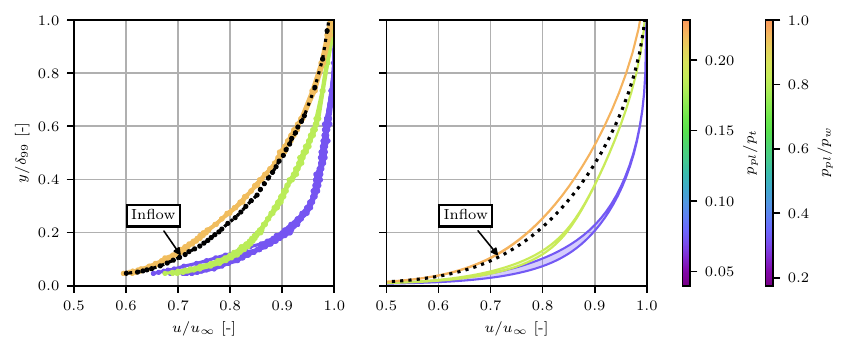}
    \begin{subfigure}[t!]{0.32\textwidth}
        \centering
        \caption{Experiments}
        \label{fig:profilesDownHRspana}
    \end{subfigure}
    \begin{subfigure}[t!]{0.40\textwidth}
        \centering
        \caption{Simulations}
        \label{fig:profilesDownHRspanb}
    \end{subfigure}
    \caption{Variations in the boundary-layer profiles \SI{15}{\mm} downstream of the bleed region along the span}
    \label{fig:profilesDownHRspan}
\end{figure*}

For a more detailed validation, multiple boundary-layer profiles at different span wise positions are acquired numerically and experimentally, and shown in Fig.~\ref{fig:profilesDownHRspan}. Fig.~\ref{fig:profilesDownHRspanb} visualizes the envelopes of profiles dependent on the pressure ratio. A significant difference between high and low-pressure ratios is apparent. For high-pressure ratios, which correspond to low bleed rates, the profiles are found to be homogeneous in the simulations. The hole geometry does not cause any three-dimensional disturbances. With lower plenum pressures, the bleed rate increases, as well as the variability in the boundary-layer profiles along the span, which is mainly found in the lower \SI{20}{\%} of the boundary layer. Further increase of the bleed rate by lowering the plenum pressure enhances the variability as the strength of the barrier shocks and expansion fans induced by the holes increase.

\begin{figure}[!b]
    \centering
    \begin{subfigure}[t!]{0.45\textwidth}
        \centering
        \makebox[\textwidth][c]{\includegraphics[trim={0mm 5mm 80mm 15mm}, clip]{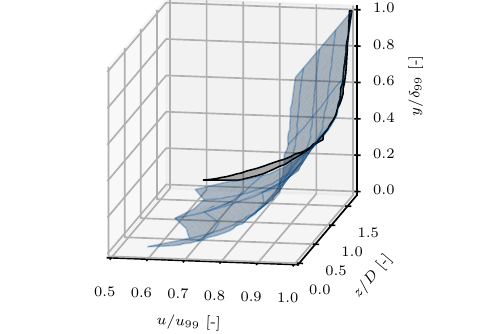}}
        \caption{Experiments}
        \label{fig:profiles3Da}
    \end{subfigure}
    \begin{subfigure}[t!]{0.45\textwidth}
        \centering
        \makebox[\textwidth][c]{\includegraphics[trim={80mm 5mm 0mm 15mm}, clip]{ASTprofiles3DCFDEXP.pdf}}
        \caption{Simulations}
        \label{fig:profiles3Db}
    \end{subfigure}
    \caption{Illustration of the velocity profiles along the span for experiments ($p_{pl}/p_t=0.072$) and simulations ($p_{pl}/p_t=0.073$)}
    \label{fig:profiles3D}
\end{figure}

The experiments mirror the same trend, as visible in Fig.~\ref{fig:profilesDownHRspana}. For the highest pressure ratio, the measurements at three different span wise positions vary only marginally. For a lower pressure ratio, the variation is found to be larger between the three positions. For the lowest pressure ratio, the measurements were conducted at eight different positions, showing the largest variability. All these measurements are shown in comparison to the simulations in Fig.~\ref{fig:profiles3D}, showing a periodic variation in the boundary-layer profiles caused by the periodic pattern of the holes.

\subsection{Subsonic boundary-layer bleeding}

After investigating supersonic boundary-layer bleeding, we focus on subsonic flows since we found them downstream of normal shock-wave/boundary-layer interactions. Therefore, experiments were conducted in the same working section using the convergent-divergent nozzle in an entirely subsonic regime, resulting in a free-stream Mach number of $M=0.5$. The subsonic experiments were limited to the $D=$ \SI{0.5}{\mm} as simulations have shown no significant variations of the boundary-layer profiles along the span. Moreover, the lower porosity level results in a lower bleed mass flow rate, which facilitates keeping the inflow conditions constant. Since the subsonic bleed influences the flow in the upstream direction, high bleed rates may result in fuller inflow profiles or an increase of the external Mach number.

\begin{figure*}[!b]
    \centering
    \makebox[\textwidth][c]{\includegraphics[trim={0mm 24mm 0mm 17mm}, clip]{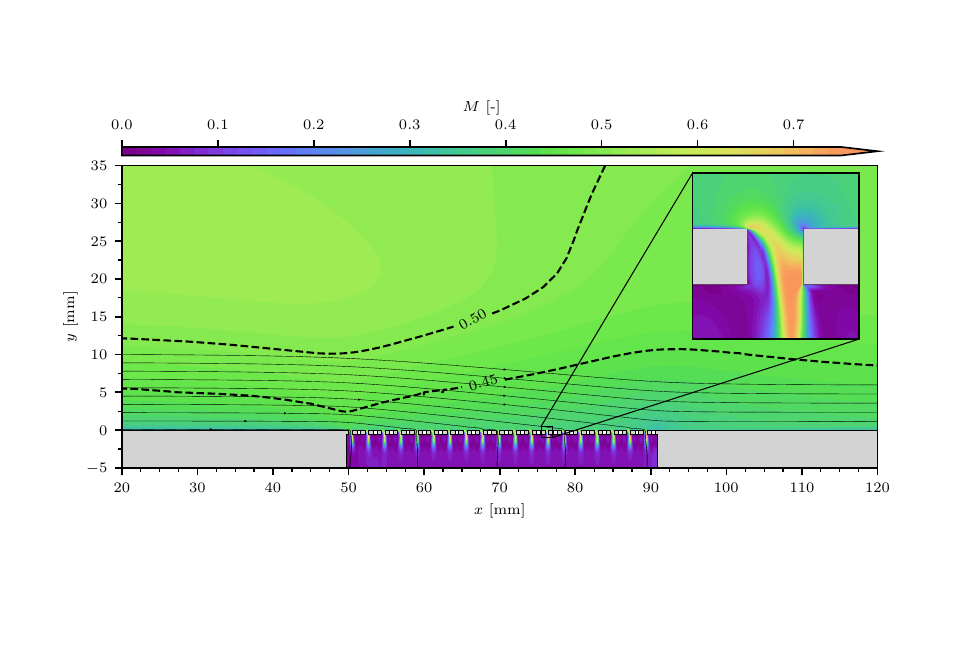}}
    \caption{Mach contour field for plate HR ($p_{pl}/p_t=0.651$); white areas illustrate holes of the second column}
    \label{fig:contourSub}
\end{figure*}

The flow topology around a subsonic bleed is illustrated in Fig.~\ref{fig:contourSub}. Like a supersonic bleed, the flow is redirected towards the wall, creating a diverging stream tube. Contrary to supersonic flows, the result is not an acceleration but a deceleration of the flow, as highlighted by the contour lines for the Mach number $M=0.45$ and $M=0.5$. Since the observed case in an internal flow, the diffuser effect is even more pronounced as the stream tube is limited in size. However, simulations without a top wall limiting the domain have shown similar results. Also, a strong upstream influence is noted in contrary to the supersonic bleed, resulting in an acceleration of the flow upstream of the bleed. The zoomed view gives an impression of the flow topology inside the holes. In this case, the flow is fully subsonic. In comparison to Fig.~\ref{fig:contour05}, the separated area in the front of the hole is smaller. Since no barrier shock exists, the losses are smaller compared to the supersonic case, and the effectiveness of the bleed in increasing the flow momentum in the wall vicinity is assumed to be higher.

\begin{figure*}[t!]
    \centering
    \includegraphics[trim={0mm 2mm 0mm 2mm}, clip]{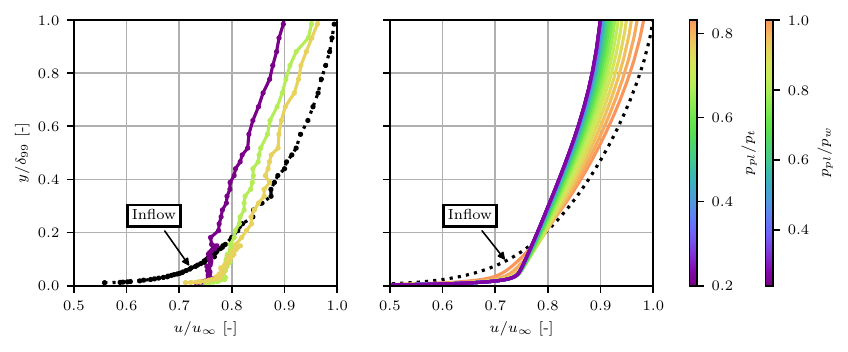}
    \begin{subfigure}[t!]{0.32\textwidth}
        \centering
        \caption{Experiments}
        \label{fig:profilesSuba}
    \end{subfigure}
    \begin{subfigure}[t!]{0.40\textwidth}
        \centering
        \caption{Simulations}
        \label{fig:profilesSubb}
    \end{subfigure}
    \caption{Boundary layer profiles \SI{15}{\mm} downstream of the bleed region for subsonic conditions}
    \label{fig:profilesSub}
\end{figure*}

In the next step, we focus on the effect of the porous bleed on the boundary-layer profiles. Fig.~\ref{fig:profilesSub} shows the state of the boundary layer upstream and downstream of the porous bleed acquired from the experiments and the simulations. Again, the profiles are normalized by the inflow boundary-layer thickness and the free-stream velocity. The simulations (see Fig.~\ref{fig:profilesSubb}) demonstrate the effect of the bleed. Even low bleed rates lead to an increase in the momentum in the lower \SI{20}{\%} of the boundary layer. Lowering the pressure ratio, and in turn, the bleed rate, results in a saturation of the effect. Below a pressure ratio of $p_{pl}/p_t \approx 0.7$, the maximum achievable improvement is achieved. On the contrary, there is still a change in the effect for the outer boundary layer notable. A further decrease in the pressure ratio increases the bleed rate and consequently reduces the external flow velocity as more mass is removed and the bleed generates a larger divergent stream tube.

The experimentally measured boundary-layer profiles are illustrated in Fig.~\ref{fig:profilesSuba}. The findings are similar to the simulations, even though the velocity in the vicinity of the wall is found to be smaller for low-pressure ratios. This deviation is assumed to be a consequence of the experimental methodology.
As the bleed has an upstream influence, high bleed rates lead to higher inflow mass flow rates. Moreover, the inflow profile becomes asymmetric, meaning the free-stream velocity is higher in the lower part of the wind tunnel where the suction occurs compared to the upper, uncontrolled flow. The missing momentum in the upper part increases the diffuser effect, and thus, even the velocity in the boundary layer is found to be smaller. However, the inflow profiles measured upstream of the bleed region were equal independently of the bleed rate. In contrast, the numerical inlet was defined to be constant outside the boundary layer. Thus, the overall mass flowing through the working section might be higher as the momentum in the upper part is larger.

\subsection{Shock/boundary-layer interaction control}

\begin{figure}[!tbh]
    \centering
    \begin{subfigure}[t!]{1.00\textwidth}
        \centering
        \makebox[\textwidth][c]{\includegraphics[trim={0mm 65mm 0mm 25mm}, clip]{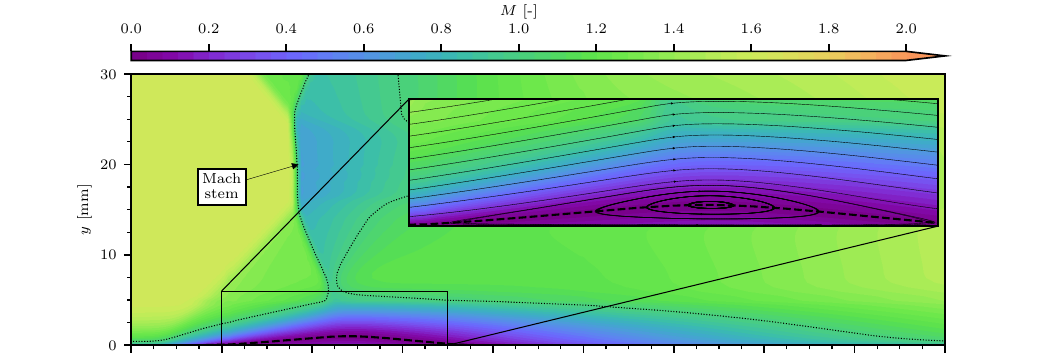}}
        \caption{Uncontrolled case; dashed line highlights separated area}
        \label{fig:sblitopa}
    \end{subfigure}
    \begin{subfigure}[t!]{1.00\textwidth}
        \centering
        \makebox[\textwidth][c]{\includegraphics[trim={0mm 9mm 0mm 93mm}, clip]{ASTtopologySBLI.pdf}}
        \caption{Controlled case with plate AR ($p_{pl}/p_t = 0.116$)}
        \label{fig:sblitopb}
    \end{subfigure}
    \caption{Mach contours for shock-wave/boundary-layer interaction in controlled and uncontrolled case; dotted lines highlight sonic line}
    \label{fig:sblitop}
\end{figure}

In the last step, the control of a shock-wave/boundary-layer interaction is investigated. A shock generator is mounted with an angle-of-attack of $\alpha=$ \SI{9.5}{\degree} in the working section, as visible in Fig.~\ref{fig:experiment}. Simulations were conducted in the Q2D setup with symmetry planes on the front and back of the domain, as well as three-dimensionally by simulating the WTHS with the HR plate. In the Q2D numerical domain, the angle-of-attack has to be increased by \SI{1.0}{\degree} to obtain the same location of the incident shock as the three-dimensionality of the flow induced by the corner flow leads to a more intense shock.

The flow field around the incident shock is illustrated in Fig.~\ref{fig:sblitop} for the Q2D simulations. In the top (Fig.~\ref{fig:sblitopa}), the uncontrolled case is shown. The deflection angle of the incident shock is too high for a regular reflection of the shock, leading to a so-called Mach-reflection. Thus, a normal shock (Mach stem) is found in the vicinity of the wall. The adverse pressure gradient of the shock is sufficiently high to make the flow separate, resulting in a separation bubble with its length being approximately the size of the bleed region.

The successfully controlled interaction is shown in Fig.~\ref{fig:sblitopb}, demonstrating the effect of the porous bleed. The flow separation is significantly mitigated, resulting in the elimination of the Lambda shock foot and a downstream movement of the shock foot. Since the porous bleed induces a bending of the flow towards the wall, the deflection angle along the shock, and hence the shock intensity decreases. Especially downstream of the incident shock, a strong aspiration flow is generated, resulting in a seemingly regular shock reflection. However, a small Mach-stem is found and decelerates the flow to subsonic conditions in the wall vicinity.

The zoom-views visualize the flow topology inside the holes upstream and downstream of the shock. On the right side, the flow topology is identical to the bleed in the purely supersonic case. The expansion fan at the hole front and the barrier shock are notable. In this area, the momentum in boundary layer is increased to decrease the upstream influence of the adverse pressure gradient. In contrast, the flow inside the holes downstream of the shock is identical to a subsonic porous bleed. The flow around the hole is purely subsonic and accelerated to supersonic conditions at the hole inlet because of the low-pressure ratio. The separated region in the front of the hole is found to be significantly smaller than upstream of the shock, resulting in higher bleed rates. The presence of different flow topologies upstream and downstream of the incident shock is essential for the derivation of an adequate porous bleed model since both super- and subsonic regimes need to be considered to predict the effect of the porous bleed on the shock-wave/boundary-layer interaction.

\begin{figure*}[!bht]
    \centering
    \makebox[\textwidth][c]{\includegraphics[trim={0mm 2mm 0mm 2mm}, clip]{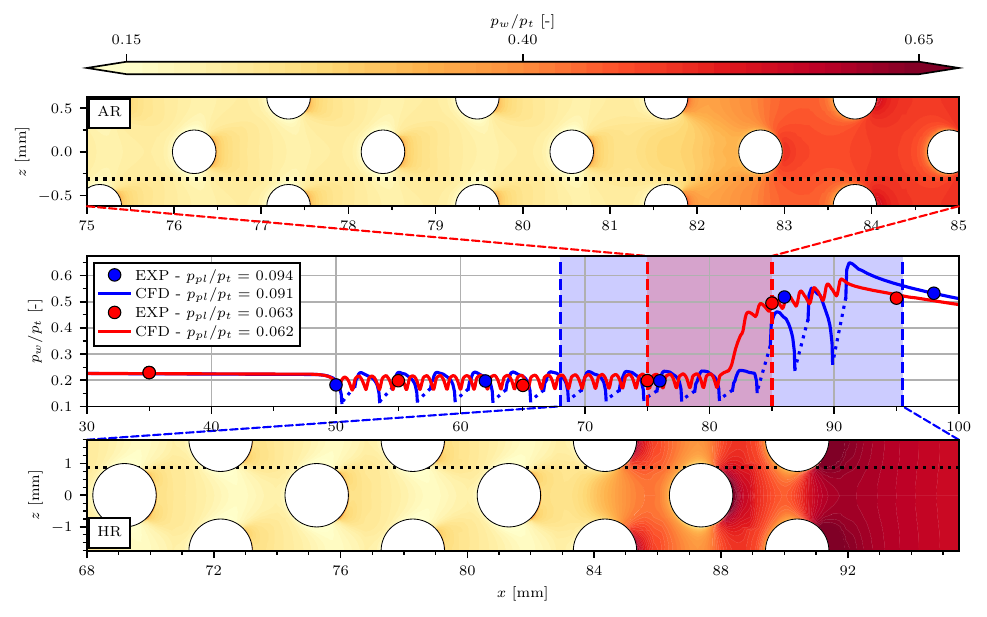}}
    \caption{Static wall pressure along the bleed region}
    \label{fig:pressureSBLI}
\end{figure*}

The trend of the wall pressure is shown for both plates in Fig.~\ref{fig:pressureSBLI}. Again, the numerical pressure contour fields are visualized at the top and the bottom while the curves are compared to the acquired data from the pressure taps. For both plates, the pressure fields focus on the region around the shock foot to illustrate the pressure rise. A different location of the shock foot is apparent, which is caused by the different porosity levels, resulting in a higher bleed rate in the case of plate HR.
At the center, the curves along the bleed region are shown. Here again, the shock foot is found to be located further upstream for plate AR (red), as seen by the pressure rise around $x=$ \SI{82}{\mm}. Further downstream, the pressure continues to increase steadily until the end of the bleed region, terminated by the trailing shock. The same trend is apparent for plate HR, with the difference of a further downstream located pressure rise. Moreover, the experimentally acquired pressure values fit very well with the simulations. 

\begin{figure}[!hbt]
    \centering
    \begin{subfigure}[t!]{1.00\textwidth}
        \centering
        \makebox[\textwidth][c]{\includegraphics[trim={0mm 71mm 0mm 30mm}, clip]{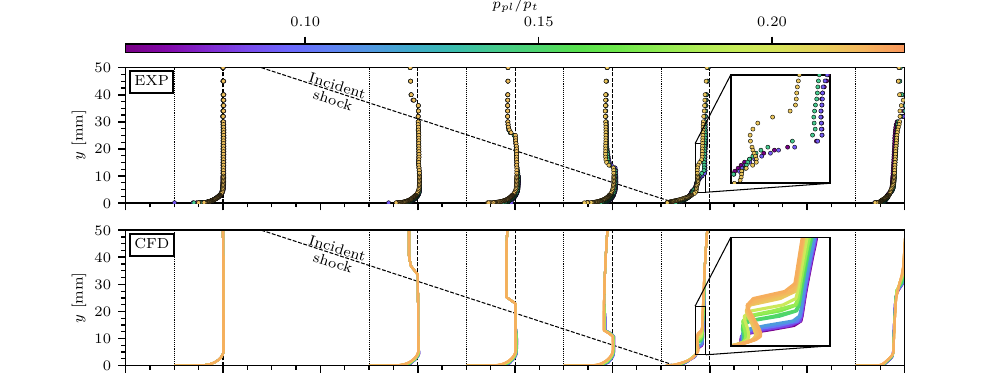}}
        \caption{Unscaled velocity profiles}
        \label{fig:sbliprofilesa}
    \end{subfigure}
    \begin{subfigure}[t!]{1.00\textwidth}
        \centering
        \makebox[\textwidth][c]{\includegraphics[trim={0mm 9mm 0mm 95mm}, clip]{ASTsbliProfilesAR.pdf}}
        \caption{Velocity profiles inside the boundary layer}
        \label{fig:sbliprofilesb}
    \end{subfigure}
    \caption{Velocity profiles along the bleed region for plate AR}
    \label{fig:sbliprofiles}
\end{figure}

For the comparison of experiments and simulations, velocity profiles were acquired along the bleed region. As described in Sec.~\ref{sec:sup3d}, the large size of the holes of the plate HR results in strong variations of the boundary-layer profiles along the span. For this reason, LDV measurements have been only performed on plate AR to keep the number of required measurements low. All the measurements, as well as the numerically extracted profiles, are shown in Fig.~\ref{fig:sbliprofiles}. At the top, the unscaled profiles are illustrated to evaluate the conformity of the overall flow field for the Q2D simulations and the experiments. In both cases, the incident shock has the same shock angle, as seen by the knee in the profiles in Fig.~\ref{fig:sbliprofilesa} independent of the bleed rate. In contrast, the shock reflection is found to be affected by the porous bleed. The lower the pressure ratio and the higher the bleed rate, the further the downstream movement of the shock. Hence, the reflected shock is located closer to the wall for the further downstream positions. The zoom-view demonstrates the effect of the bleed rate on the velocity profiles. The agreement between experiments and simulations is very good.

For the observation of the near-wall region inside the boundary layer, the profiles are scaled by the incoming boundary-layer thickness (see Fig.~\ref{fig:sbliprofilesb}). The profiles upstream of the incident shock, which is located between $x=$ \SIlist{80; 85}{\mm}, show the increase of the momentum inside the boundary layer dependent of the bleed rate. The higher the bleed rate, the fuller the boundary-layer profile, as visible in the zoom-view on the left. Please note that the knees in the numerical profiles are caused by the relative position to the bleed holes and the resulting passing of shocks and expansion waves. In the experiments, these effects are more smeared because of the size of the measurement volume.

Downstream of the shock, artifacts of the supersonic control are notable. The higher the momentum in the boundary layer upstream of the shock the higher the momentum downstream of the shock. However, a higher momentum downstream of the shock is expected to result in a lower bleed efficiency as the flow separation inside the holes increases, and hence, the vena contracta area decreases. Since the adverse pressure gradient along the plate below the shock foot leads to lower pressure ratios $p_{pl}/p_w$, the relative difference between the cases becomes smaller, and the bleed works close to, or under choked conditions ($p_{pl}/p_w \leq 0.528$). Thus, a lower bleed efficiency results in lower bleed rates, and a lower increase of flow momentum near the wall. Consequently, the differences between the cases are found to be smaller than upstream of the shock, as illustrated in the zoom-view on the right. Downstream of the bleed region, the variations in the boundary-layer profiles are even smaller compared to the differences upstream of the shock.

\begin{figure}[!tbh]
    \centering
    \begin{subfigure}[t!]{1.00\textwidth}
        \centering
        \makebox[\textwidth][c]{\includegraphics[trim={0mm 96mm 0mm 24mm}, clip]{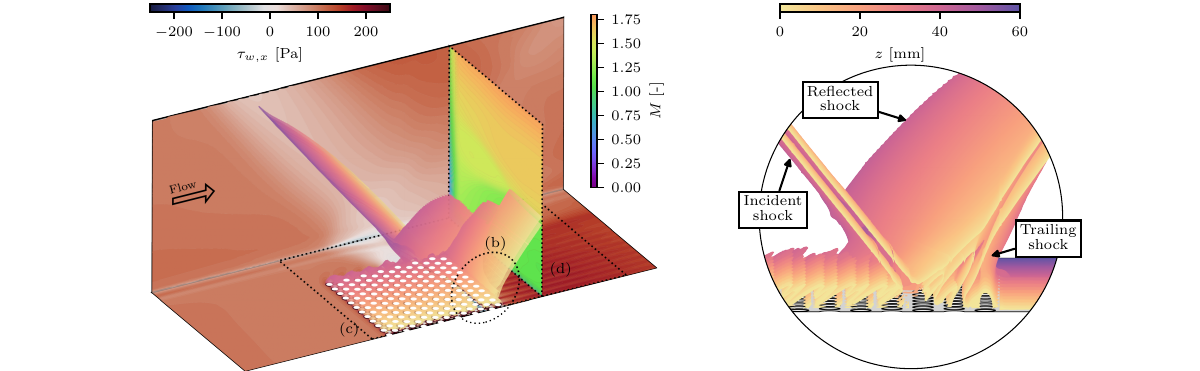}}
    \end{subfigure}
    \begin{subfigure}[t!]{0.64\textwidth}
        \centering
        \caption{Flow field inside the wind tunnel}
        \label{fig:shockTunnela}
    \end{subfigure}
    \begin{subfigure}[t!]{0.35\textwidth}
        \centering
        \caption{Density gradient iso-contour}
        \label{fig:shockTunnelb}
    \end{subfigure}
    \begin{subfigure}[t!]{1.00\textwidth}
        \centering
        \makebox[\textwidth][c]{\includegraphics[trim={0mm 12mm 0mm 104mm}, clip]{ASTflowST16.pdf}}
    \end{subfigure}
    \begin{subfigure}[t!]{0.64\textwidth}
        \centering
        \caption{Wall shear stress around the bleed region}
        \label{fig:shockTunnelc}
    \end{subfigure}
    \begin{subfigure}[t!]{0.35\textwidth}
        \centering
        \caption{Flow field downstream of the bleed region}
        \label{fig:shockTunneld}
    \end{subfigure}
    \caption{Flow topology with shock-wave/boundary-layer interaction inside the wind tunnel from simulation ($p_{pl}/p_t=0.095$)}
    \label{fig:shockTunnel}
\end{figure}

In the last step, the WTHS simulations are compared to the flow visualization for the plate HR. Fig.~\ref{fig:shockTunnel} gives an overview of the flow topology inside the wind tunnel. In Fig.~\ref{fig:shockTunnela}, the wall shear stress as well as the flow field downstream of the bleed region are shown. Moreover, an iso-contour of the density gradient is added to visualize the incident as well as the reflected shock. Directly at the first glance, a significant three-dimensionality of the flow is observable. The curvature of the shock is used as a measure to evaluate the variation of the flow along the span. Therefore, the iso-contour of the density gradient is closer observed in Fig.~\ref{fig:shockTunnelb}. The darker the color of the iso-contour, the closer the location with regard to the wind tunnel side wall. The incident shock has only a small variation along the span and is only slightly curved in the upstream direction towards the side wall. On the contrary, the reflected shock is strongly curved. The low flow momentum of the corner flow increases the upstream influence of the incident shock, moving the shock foot and the reflected shock upstream. Interestingly, the reflected shock merges with the trailing shock at the center of the working section.

The footprint of the shock is visible in Fig.~\ref{fig:shockTunnelc} which shows the wall shear stress. The positive pressure gradient induces a local flow separation, visualized by the blue color range and surrounded by the dashed line. The size of the separation is limited to the area between the holes, as the suction leads to a fast reattachment of the flow. At the center of the working section, the shock can be located at $x\approx$ \SI{83}{\mm} since the flow starts to separate even between the holes at this station. Along the span, the incident flow separation moves upstream. Alongside the bleed region, a small area of separated flow is apparent, as well as a separation of the corner flow. However, the flow separation does not cover the entire span of the wind tunnel as the bleed energizes the flow at the center. The streamlines demonstrate how the suction confines the width of the stream tube. Similar effects were found in earlier investigations with vortex generators controlling the center flow~\cite{Titchener2011a, Titchener2013}.

The flow field on a slice downstream of the bleed region is shown in Fig.~\ref{fig:shockTunneld}. In the upper part, the Mach number is found to be higher as the reflected shock has not passed the flow at this location, while the Mach number is significantly reduced in the lower part of the working section. The iso-line of the density gradient pinpoints the approximate position of the reflected shock. Here again, the merging of trailing and reflected shock is observed in the center of the working section. With further distance from the center, the two shocks divide. Near the side walls, the intensity of the shocks decreases as they merge with the expansion waves caused by the bleed and the rear edge of the shock generator. Particularly the expansion fan caused by the shock generator results in a smearing of the density gradient and leads to a re-acceleration of the flow.

Furthermore, the sonic height is extracted and highlighted by the dotted line. Its high amount at the side walls near the corner illustrates the thickening of the boundary layer caused by the adverse pressure gradient of the shock. Thus, the sonic height is larger at the top before passing the shock and increases against the corner. In the area downstream of the bleed, the sonic height is smaller as the suction increases the momentum inside the boundary layer. The lowest sonic height is found around $x=$ \SI{30}{\mm} with its value slightly increasing towards the center of the working section. As the shock position is further upstream towards the side wall, the area with a subsonic working regime of the bleed is larger. Hence, the effect of the flow control is increased in this area and the momentum is higher in the wall vicinity. However, no significant effect of the bleed on the corner flow is apparent. The presence of an expansion fan caused by the rear edge of the shock generator leads to a reattachment of the flow and counteracts the negative effect of the porous bleed.

\begin{figure}[!hbt]
    \centering
    \begin{subfigure}[t!]{1.00\textwidth}
        \centering
        \makebox[\textwidth][c]{\includegraphics[trim={0mm 49mm 0mm 2mm}, clip]{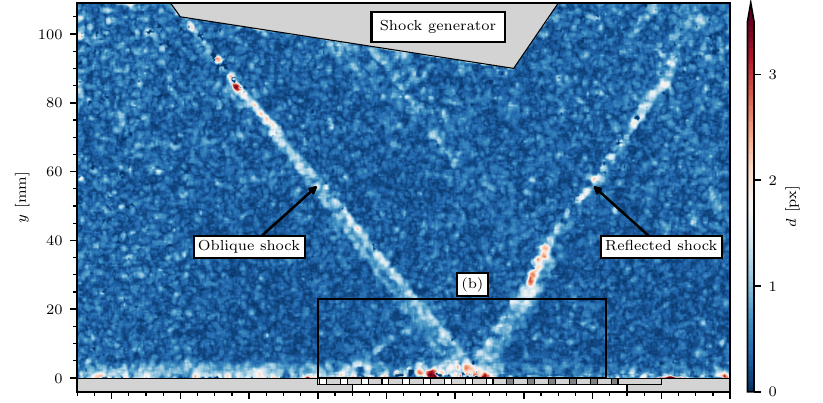}}
        \caption{View on the working section including the shock generator}
        \label{fig:sblibosa}
    \end{subfigure}
    \begin{subfigure}[t!]{1.00\textwidth}
        \centering
        \makebox[\textwidth][c]{\includegraphics[trim={0mm 2mm 0mm 77mm}, clip]{ASTbosSBLI.pdf}}
        \caption{Zoom-view on the shock-wave/boundary-layer interaction}
        \label{fig:sblibosb}
    \end{subfigure}
    \caption{BOS visualization of the shock-wave/boundary-layer interaction control for $p_{pl}/p_t = 0.094$}
    \label{fig:sblibos}
\end{figure}

BOS visualizations of the experiments with the plate HR are shown in Fig.~\ref{fig:sblibos}. At the top, the flow field is illustrated from a distance, showing almost the entire working section including the shock generator. The generated oblique shock is seemingly regularly reflected at the bottom wall. Faintly visible is the expansion fan created by the bleed region, as well as the expansion fan generated by the rear edge of the shock generator. A closer look at the interaction is shown in Fig.~\ref{fig:sblibosb}, where the camera was positioned closer to the working section focusing on the impingement point of the shock, revealing more information about the shock structure. A distinct incident shock is visible with its impingement point located at $x\approx$ \SI{85}{\mm}. The reflected shock is divided in two distinct parts and smeared between. One part of the shock has a significantly lower shock angle and is located further downstream, where it merges with the trailing shock. The second distinct part is located further upstream and forms a shock wave located with an offset to the merged shock. Since the BOS visualization is an integration of the density gradient along the visual axis of the camera, it is expected that the first part of the reflected shock is located close to the center of the wind tunnel, while the second part is found near to the side walls. Thus, the structure of the shock system is in line with the numerical findings.

\section{Conclusion}

This paper studied experimentally and numerically the control of boundary layer by porous bleed systems. We have presented the effect of the porous bleed on a shock-wave/boundary-layer interaction after simplifying the complex problem into two separate problems: super- and subsonic boundary-layer control by porous bleed. The findings support the idea to separately characterize the working principle of porous bleed systems for both flow regimes as the flow topology inside the bleed holes significantly differs, resulting in different effects on the boundary layer and interaction with an incident shock.

In the supersonic flow case, we have found a good agreement between experimental and numerical results. The effect of the porous bleed on the boundary layer is well resolved by the numerical simulations, as proven by the LDV-obtained velocity fields, boundary-layer profiles, and static wall pressure measurements. Moreover, three-dimensional effects resulting in variations of the boundary-layer profiles along the span for the use of large bleed holes were proven for the first time in experiments. Furthermore, the effect on the corner flow and the impact of the corner flow on the two-dimensionality of the center flow was illustrated.

In addition, the effect of the porous bleed on a subsonic flow was determined in the same experimental and numerical setup. For the first time, one study described the working principle of porous bleeds in both super- and subsonic conditions by means of experiments and RANS simulations. Like in the supersonic regime, the boundary-layer bleeding leads to an increase of the flow momentum in the wall vicinity. However, the removal of mass results in a decrease of the flow momentum in the outer region of the boundary layer as well as of the external flow.

Finally, the shock-wave/boundary-layer interaction control was studied. Even for this complex flow topology, a very good agreement between simulations and experiments was achieved. The effect on the boundary layer along the bleed region was well replicated by the experiments. Also, three-dimensional simulations of the WTHS were performed to study the impact of the corner flow on the interaction of shock, boundary layer, and porous bleed. Like in previous studies, a reinforcement of the corner flow was observed. In both experiments and simulations, the same structure of the shock topology was found.

The findings of this study legitimate the use of RANS simulations to investigate the effect of porous bleed systems in subsonic and supersonic flows with and without shock-wave/boundary-layer interactions. As experiments are time- and cost-intensive to examine comprehensive parametric studies, numerical simulations can help to deepen the knowledge about the working principle of bleed systems to improve bleed models. Moreover, the findings endorse a separate consideration of super- and subsonic bleeds, especially with regard to bleed models that are required to work in the case of normal shock-wave/boundary-layer interactions.

\section*{Acknowledgments}
This project has received funding from the European Union's Horizon 2020 research and innovation program under grant agreement No EC grant 860909. This support has been essential in the successful completion of this project. We would also like to thank our institute for their support, as well as Francois Nicolas and Cedric Illoul for assisting with the BOS measurements.


\begin{thebibliography}{27}
\expandafter\ifx\csname natexlab\endcsname\relax\def\natexlab#1{#1}\fi
\providecommand{\url}[1]{\texttt{#1}}
\providecommand{\href}[2]{#2}
\providecommand{\path}[1]{#1}
\providecommand{\DOIprefix}{doi:}
\providecommand{\ArXivprefix}{arXiv:}
\providecommand{\URLprefix}{URL: }
\providecommand{\Pubmedprefix}{pmid:}
\providecommand{\doi}[1]{\href{http://dx.doi.org/#1}{\path{#1}}}
\providecommand{\Pubmed}[1]{\href{pmid:#1}{\path{#1}}}
\providecommand{\bibinfo}[2]{#2}
\ifx\xfnm\relax \def\xfnm[#1]{\unskip,\space#1}\fi
\bibitem[{Babinsky and Ogawa(2008)}]{Babinsky2008}
\bibinfo{author}{H.~Babinsky}, \bibinfo{author}{H.~Ogawa},
\newblock \bibinfo{title}{{SBLI control for wings and inlets}},
\newblock \bibinfo{journal}{Shock Waves} \bibinfo{volume}{18}
  (\bibinfo{year}{2008}) \bibinfo{pages}{89--96}. \URLprefix
  \url{http://link.springer.com/10.1007/s00193-008-0149-7}.
  \DOIprefix\doi{10.1007/s00193-008-0149-7}.
\bibitem[{Syberg and Hickcox(1973)}]{Syberg1973a}
\bibinfo{author}{J.~Syberg}, \bibinfo{author}{T.~E. Hickcox},
  \bibinfo{title}{{Design of a Bleed System for a Mach 3. 5 Inlet.}},
  \bibinfo{type}{Technical Report}, National Aeronautics and Space
  Administration, \bibinfo{address}{Washington, DC}, \bibinfo{year}{1973}.
\bibitem[{Harloff and Smith(1996)}]{Harloff1996}
\bibinfo{author}{G.~J. Harloff}, \bibinfo{author}{G.~E. Smith},
\newblock \bibinfo{title}{{Supersonic-inlet boundary-layer bleed flow}},
\newblock \bibinfo{journal}{AIAA Journal} \bibinfo{volume}{34}
  (\bibinfo{year}{1996}) \bibinfo{pages}{778--785}. \URLprefix
  \url{https://arc.aiaa.org/doi/10.2514/3.13140}.
  \DOIprefix\doi{10.2514/3.13140}.
\bibitem[{Shih(2008)}]{Shih2008}
\bibinfo{author}{T.~Shih},
\newblock \bibinfo{title}{{Control of Shock-Wave/Bound-Layer Interactions by
  Bleed}},
\newblock \bibinfo{journal}{International Journal of Fluid Machinery and
  Systems} \bibinfo{volume}{1} (\bibinfo{year}{2008}) \bibinfo{pages}{24--32}.
  \URLprefix
  \url{http://koreascience.or.kr/journal/view.jsp?kj=OCGKEU{\&}py=2008{\&}vnc=v1n1{\&}sp=24}.
  \DOIprefix\doi{10.5293/IJFMS.2008.1.1.024}.
\bibitem[{Hamed et~al.(2011)Hamed, Manavasi, Shin, Morell, and
  Nelson}]{Hamed2011}
\bibinfo{author}{A.~Hamed}, \bibinfo{author}{S.~Manavasi},
  \bibinfo{author}{D.~Shin}, \bibinfo{author}{A.~T. Morell},
  \bibinfo{author}{C.~Nelson},
\newblock \bibinfo{title}{{Bleed Interactions in Supersonic Flow}},
\newblock \bibinfo{journal}{International Journal of Flow Control}
  \bibinfo{volume}{3} (\bibinfo{year}{2011}) \bibinfo{pages}{37--48}.
  \URLprefix
  \url{http://multi-science.atypon.com/doi/10.1260/1756-8250.3.1.37}.
  \DOIprefix\doi{10.1260/1756-8250.3.1.37}.
\bibitem[{Oorebeek et~al.(2015)Oorebeek, Babinsky, Ugolotti, Orkwis, and
  Duncan}]{Oorebeek2015}
\bibinfo{author}{J.~Oorebeek}, \bibinfo{author}{H.~Babinsky},
  \bibinfo{author}{M.~Ugolotti}, \bibinfo{author}{P.~D. Orkwis},
  \bibinfo{author}{S.~Duncan},
\newblock \bibinfo{title}{{Experimental and Computational Investigations of a
  Normal-Hole-Bled Supersonic Boundary Layer}},
\newblock \bibinfo{journal}{AIAA Journal} \bibinfo{volume}{53}
  (\bibinfo{year}{2015}) \bibinfo{pages}{3726--3736}. \URLprefix
  \url{https://arc.aiaa.org/doi/10.2514/1.J053956}.
  \DOIprefix\doi{10.2514/1.J053956}.
\bibitem[{Zhang et~al.(2020)Zhang, Zhao, Liu, and Wang}]{Zhang2020b}
\bibinfo{author}{B.-h. Zhang}, \bibinfo{author}{Y.-x. Zhao},
  \bibinfo{author}{J.~Liu}, \bibinfo{author}{Q.~Wang},
\newblock \bibinfo{title}{{Supersonic Bleed Rate Model for a Circular Orifice
  Based on Geometric Similarity}},
\newblock \bibinfo{journal}{AIAA Journal} \bibinfo{volume}{58}
  (\bibinfo{year}{2020}) \bibinfo{pages}{2486--2493}. \URLprefix
  \url{https://arc.aiaa.org/doi/10.2514/1.J058495}.
  \DOIprefix\doi{10.2514/1.J058495}.
\bibitem[{Schwartz et~al.(2023)Schwartz, Gaitonde, and Slater}]{Schwartz2023}
\bibinfo{author}{M.~J. Schwartz}, \bibinfo{author}{D.~V. Gaitonde},
  \bibinfo{author}{J.~W. Slater},
\newblock \bibinfo{title}{{Uncertainty and Sensitivity Analysis of Bleed
  Modeling in Shock/Turbulent Interactions}},
\newblock \bibinfo{journal}{Journal of Propulsion and Power}
  \bibinfo{volume}{39} (\bibinfo{year}{2023}) \bibinfo{pages}{106--120}.
  \URLprefix \url{https://arc.aiaa.org/doi/10.2514/1.B38785}.
  \DOIprefix\doi{10.2514/1.B38785}.
\bibitem[{Giehler et~al.(2023)Giehler, Grenson, and Bur}]{Giehler2023b}
\bibinfo{author}{J.~Giehler}, \bibinfo{author}{P.~Grenson},
  \bibinfo{author}{R.~Bur},
\newblock \bibinfo{title}{{Parameter Influence on Porous Bleed Performance for
  Supersonic Turbulent Flows}},
\newblock \bibinfo{journal}{Journal of Propulsion and Power}
  (\bibinfo{year}{2023}) \bibinfo{pages}{1--20}. \URLprefix
  \url{https://arc.aiaa.org/doi/10.2514/1.B39236}.
  \DOIprefix\doi{10.2514/1.B39236}.
\bibitem[{Willis et~al.(1995)Willis, Davis, and Hingst}]{Willis1995b}
\bibinfo{author}{B.~Willis}, \bibinfo{author}{D.~O. Davis},
  \bibinfo{author}{W.~R. Hingst},
\newblock \bibinfo{title}{{Flow coefficient behavior for boundary layer bleed
  holes and slots}},
\newblock in: \bibinfo{booktitle}{33rd Aerospace Sciences Meeting and Exhibit},
  \bibinfo{publisher}{American Institute of Aeronautics and Astronautics},
  \bibinfo{address}{Reston, Virginia}, \bibinfo{year}{1995}. \URLprefix
  \url{https://arc.aiaa.org/doi/10.2514/6.1995-31}.
  \DOIprefix\doi{10.2514/6.1995-31}.
\bibitem[{Willis and Davis(1996)}]{Willis1996}
\bibinfo{author}{B.~Willis}, \bibinfo{author}{D.~O. Davis},
\newblock \bibinfo{title}{{Boundary layer development downstream of a bleed
  mass flow removal region}},
\newblock in: \bibinfo{booktitle}{32nd Joint Propulsion Conference and
  Exhibit}, \bibinfo{publisher}{American Institute of Aeronautics and
  Astronautics}, \bibinfo{address}{Reston, Virginia}, \bibinfo{year}{1996}.
  \URLprefix \url{https://arc.aiaa.org/doi/10.2514/6.1996-3278}.
  \DOIprefix\doi{10.2514/6.1996-3278}.
\bibitem[{Eichorn et~al.(2013)Eichorn, Barnhart, Davis, Vyas, and
  Slater}]{Eichorn2013}
\bibinfo{author}{M.~Eichorn}, \bibinfo{author}{P.~Barnhart},
  \bibinfo{author}{D.~O. Davis}, \bibinfo{author}{M.~Vyas},
  \bibinfo{author}{J.~W. Slater},
\newblock \bibinfo{title}{{Effect of Boundary-Layer Bleed Hole Inclination
  Angle and Scaling on Flow Coefficient Behavior}},
\newblock in: \bibinfo{booktitle}{51st AIAA Aerospace Sciences Meeting
  including the New Horizons Forum and Aerospace Exposition},
  \bibinfo{publisher}{American Institute of Aeronautics and Astronautics},
  \bibinfo{address}{Reston, Virginia}, \bibinfo{year}{2013}. \URLprefix
  \url{https://arc.aiaa.org/doi/10.2514/6.2013-424}.
  \DOIprefix\doi{10.2514/6.2013-424}.
\bibitem[{Giehler et~al.(2023)Giehler, Grenson, and Bur}]{Giehler2023a}
\bibinfo{author}{J.~Giehler}, \bibinfo{author}{P.~Grenson},
  \bibinfo{author}{R.~Bur},
\newblock \bibinfo{title}{{Porous Bleed Boundary Conditions for Supersonic
  Flows With {\&} Without Shock-Boundary Layer Interaction}},
\newblock \bibinfo{journal}{Flow, Turbulence and Combustion}
  (\bibinfo{year}{2023}). \URLprefix
  \url{https://doi.org/10.1007/s10494-023-00464-9
  https://link.springer.com/10.1007/s10494-023-00464-9}.
  \DOIprefix\doi{10.1007/s10494-023-00464-9}.
\bibitem[{Willis et~al.(1995)Willis, Davis, and Hingst}]{Willis1995a}
\bibinfo{author}{B.~Willis}, \bibinfo{author}{D.~O. Davis},
  \bibinfo{author}{W.~R. Hingst},
\newblock \bibinfo{title}{{Flowfield measurements in a normal-hole-bled oblique
  shock-wave and turbulent boundary-layer interaction}},
\newblock in: \bibinfo{booktitle}{31st Joint Propulsion Conference and
  Exhibit}, \bibinfo{publisher}{American Institute of Aeronautics and
  Astronautics}, \bibinfo{address}{Reston, Virginia}, \bibinfo{year}{1995}.
  \URLprefix \url{https://arc.aiaa.org/doi/10.2514/6.1995-2885}.
  \DOIprefix\doi{10.2514/6.1995-2885}.
\bibitem[{Titchener and Babinsky(2011)}]{Titchener2011a}
\bibinfo{author}{N.~Titchener}, \bibinfo{author}{H.~Babinsky},
\newblock \bibinfo{title}{{Microvortex Generators Applied to a Flowfield
  Containing a Normal Shock Wave and Diffuser}},
\newblock \bibinfo{journal}{AIAA Journal} \bibinfo{volume}{49}
  (\bibinfo{year}{2011}) \bibinfo{pages}{1046--1056}. \URLprefix
  \url{https://arc.aiaa.org/doi/10.2514/1.J050760}.
  \DOIprefix\doi{10.2514/1.J050760}.
\bibitem[{Titchener and Babinsky(2013)}]{Titchener2013}
\bibinfo{author}{N.~Titchener}, \bibinfo{author}{H.~Babinsky},
\newblock \bibinfo{title}{{Shock Wave/Boundary-Layer Interaction Control Using
  a Combination of Vortex Generators and Bleed}},
\newblock \bibinfo{journal}{AIAA Journal} \bibinfo{volume}{51}
  (\bibinfo{year}{2013}) \bibinfo{pages}{1221--1233}. \URLprefix
  \url{http://arc.aiaa.org/doi/10.2514/1.J052079}.
  \DOIprefix\doi{10.2514/1.J052079}.
\bibitem[{Durst et~al.(1977)Durst, Melling, Whitelaw, and Wang}]{Durst1976}
\bibinfo{author}{F.~Durst}, \bibinfo{author}{A.~Melling},
  \bibinfo{author}{J.~H. Whitelaw}, \bibinfo{author}{C.~P. Wang},
\newblock \bibinfo{title}{{Principles and Practice of Laser-Doppler
  Anemometry}},
\newblock \bibinfo{journal}{Journal of Applied Mechanics} \bibinfo{volume}{44}
  (\bibinfo{year}{1977}) \bibinfo{pages}{518--518}. \URLprefix
  \url{https://asmedigitalcollection.asme.org/appliedmechanics/article/44/3/518/385614/Principles-and-Practice-of-LaserDoppler-Anemometry}.
  \DOIprefix\doi{10.1115/1.3424128}.
\bibitem[{Nicolas et~al.(2016)Nicolas, Todoroff, Plyer, {Le Besnerais}, Donjat,
  Micheli, Champagnat, Cornic, and {Le Sant}}]{Nicolas2016}
\bibinfo{author}{F.~Nicolas}, \bibinfo{author}{V.~Todoroff},
  \bibinfo{author}{A.~Plyer}, \bibinfo{author}{G.~{Le Besnerais}},
  \bibinfo{author}{D.~Donjat}, \bibinfo{author}{F.~Micheli},
  \bibinfo{author}{F.~Champagnat}, \bibinfo{author}{P.~Cornic},
  \bibinfo{author}{Y.~{Le Sant}},
\newblock \bibinfo{title}{{A direct approach for instantaneous 3D density field
  reconstruction from background-oriented schlieren (BOS) measurements}},
\newblock \bibinfo{journal}{Experiments in Fluids} \bibinfo{volume}{57}
  (\bibinfo{year}{2016}) \bibinfo{pages}{13}. \URLprefix
  \url{http://link.springer.com/10.1007/s00348-015-2100-x}.
  \DOIprefix\doi{10.1007/s00348-015-2100-x}.
\bibitem[{Champagnat et~al.(2011)Champagnat, Plyer, {Le Besnerais}, Leclaire,
  Davoust, and {Le Sant}}]{Champagnat2011}
\bibinfo{author}{F.~Champagnat}, \bibinfo{author}{A.~Plyer},
  \bibinfo{author}{G.~{Le Besnerais}}, \bibinfo{author}{B.~Leclaire},
  \bibinfo{author}{S.~Davoust}, \bibinfo{author}{Y.~{Le Sant}},
\newblock \bibinfo{title}{{Fast and accurate PIV computation using highly
  parallel iterative correlation maximization}},
\newblock \bibinfo{journal}{Experiments in Fluids} \bibinfo{volume}{50}
  (\bibinfo{year}{2011}) \bibinfo{pages}{1169--1182}.
  \DOIprefix\doi{10.1007/s00348-011-1054-x}.
\bibitem[{Benoit et~al.(2015)Benoit, P{\'{e}}ron, and Landier}]{Benoit2015}
\bibinfo{author}{C.~Benoit}, \bibinfo{author}{S.~P{\'{e}}ron},
  \bibinfo{author}{S.~Landier},
\newblock \bibinfo{title}{{Cassiopee: A CFD pre- and post-processing tool}},
\newblock \bibinfo{journal}{Aerospace Science and Technology}
  \bibinfo{volume}{45} (\bibinfo{year}{2015}) \bibinfo{pages}{272--283}.
  \URLprefix \url{http://dx.doi.org/10.1016/j.ast.2015.05.023
  https://linkinghub.elsevier.com/retrieve/pii/S1270963815001777}.
  \DOIprefix\doi{10.1016/j.ast.2015.05.023}.
\bibitem[{Cambier et~al.(2013)Cambier, Heib, and Plot}]{Cambier2013}
\bibinfo{author}{L.~Cambier}, \bibinfo{author}{S.~Heib},
  \bibinfo{author}{S.~Plot},
\newblock \bibinfo{title}{{The Onera elsA CFD software: input from research and
  feedback from industry}},
\newblock \bibinfo{journal}{Mechanics {\&} Industry} \bibinfo{volume}{14}
  (\bibinfo{year}{2013}) \bibinfo{pages}{159--174}. \URLprefix
  \url{http://www.mechanics-industry.org/10.1051/meca/2013056}.
  \DOIprefix\doi{10.1051/meca/2013056}.
\bibitem[{Spalart(2000)}]{Spalart2000}
\bibinfo{author}{P.~Spalart},
\newblock \bibinfo{title}{{Strategies for turbulence modelling and
  simulations}},
\newblock \bibinfo{journal}{International Journal of Heat and Fluid Flow}
  \bibinfo{volume}{21} (\bibinfo{year}{2000}) \bibinfo{pages}{252--263}.
  \URLprefix
  \url{https://linkinghub.elsevier.com/retrieve/pii/S0142727X00000072}.
  \DOIprefix\doi{10.1016/S0142-727X(00)00007-2}.
\bibitem[{Wukie et~al.(2015)Wukie, Orkwis, Turner, and Duncan}]{Wukie2015}
\bibinfo{author}{N.~A. Wukie}, \bibinfo{author}{P.~D. Orkwis},
  \bibinfo{author}{M.~G. Turner}, \bibinfo{author}{S.~Duncan},
\newblock \bibinfo{title}{{Simulations and Models for Aspiration in a
  Supersonic Flow Using OVERFLOW}},
\newblock \bibinfo{journal}{AIAA Journal} \bibinfo{volume}{53}
  (\bibinfo{year}{2015}) \bibinfo{pages}{2052--2056}. \URLprefix
  \url{https://arc.aiaa.org/doi/10.2514/1.J053214}.
  \DOIprefix\doi{10.2514/1.J053214}.
\bibitem[{Giehler et~al.(2023)Giehler, Grenson, and Bur}]{Giehler2023c}
\bibinfo{author}{J.~Giehler}, \bibinfo{author}{P.~Grenson},
  \bibinfo{author}{R.~Bur},
\newblock \bibinfo{title}{{A New Approach of Using Porous Bleed Boundary
  Conditions - Application of Local Porosity}},
\newblock in: \bibinfo{editor}{A.~Dillmann}, \bibinfo{editor}{G.~Heller},
  \bibinfo{editor}{E.~Kr{\"{a}}mer}, \bibinfo{editor}{C.~Wagner},
  \bibinfo{editor}{J.~Weiss} (Eds.), \bibinfo{booktitle}{New Results in
  Numerical and Experimental Fluid Mechanics XIV}, \bibinfo{publisher}{Springer
  Nature Switzerland}, \bibinfo{address}{Cham}, \bibinfo{year}{2023}, pp.
  \bibinfo{pages}{361--371}. \URLprefix
  \url{https://link.springer.com/10.1007/978-3-031-40482-5{\_}34}.
  \DOIprefix\doi{10.1007/978-3-031-40482-5_34}.
\bibitem[{Babinsky and Harvey(2011)}]{Babinsky2011}
\bibinfo{author}{H.~Babinsky}, \bibinfo{author}{J.~K. Harvey},
  \bibinfo{title}{{Shock Wave–Boundary-Layer Interactions}},
  \bibinfo{publisher}{Cambridge University Press},
  \bibinfo{address}{Cambridge}, \bibinfo{year}{2011}. \URLprefix
  \url{http://ebooks.cambridge.org/ref/id/CBO9780511842757}.
  \DOIprefix\doi{10.1017/CBO9780511842757}.
\bibitem[{Oorebeek and Babinsky(2013)}]{Oorebeek2013}
\bibinfo{author}{J.~Oorebeek}, \bibinfo{author}{H.~Babinsky},
\newblock \bibinfo{title}{{Flow physics of a normal-hole bled supersonic
  turbulent boundary layer}},
\newblock in: \bibinfo{booktitle}{51st AIAA Aerospace Sciences Meeting
  including the New Horizons Forum and Aerospace Exposition},
  \bibinfo{publisher}{American Institute of Aeronautics and Astronautics},
  \bibinfo{address}{Reston, Virginia}, \bibinfo{year}{2013}. \URLprefix
  \url{https://arc.aiaa.org/doi/10.2514/6.2013-526}.
  \DOIprefix\doi{10.2514/6.2013-526}.
\bibitem[{Bruce et~al.(2011)Bruce, Burton, Titchener, and Babinsky}]{Bruce2011}
\bibinfo{author}{P.~J.~K. Bruce}, \bibinfo{author}{D.~M.~F. Burton},
  \bibinfo{author}{N.~A. Titchener}, \bibinfo{author}{H.~Babinsky},
\newblock \bibinfo{title}{{Corner effect and separation in transonic channel
  flows}},
\newblock \bibinfo{journal}{Journal of Fluid Mechanics} \bibinfo{volume}{679}
  (\bibinfo{year}{2011}) \bibinfo{pages}{247--262}. \URLprefix
  \url{https://www.cambridge.org/core/product/identifier/S0022112011001352/type/journal{\_}article}.
  \DOIprefix\doi{10.1017/jfm.2011.135}.

\end{thebibliography}
\end{document}